\documentclass[sigconf,anonymous=False,review=False]{acmart}
\copyrightyear{2022}
\acmYear{2022}
\setcopyright{acmlicensed}\acmConference[WWW '22]{Proceedings of the ACM Web Conference 2022}{April 25--29, 2022}{Virtual Event, Lyon, France}
\acmBooktitle{Proceedings of the ACM Web Conference 2022 (WWW '22), April 25--29, 2022, Virtual Event, Lyon, France}
\acmPrice{15.00}
\acmDOI{10.1145/3485447.3511957}
\acmISBN{978-1-4503-9096-5/22/04}

\usepackage{pifont}
\usepackage{booktabs} 
\usepackage{amsmath}
\usepackage{multirow}
\usepackage{flushend}
\usepackage{enumitem}
\usepackage[T1]{fontenc}
\usepackage[utf8]{inputenc}
\usepackage{algorithmic}
\usepackage[linesnumbered,ruled,vlined]{algorithm2e}
\usepackage[switch]{lineno}
\usepackage{eqparbox}
\usepackage[nopar]{lipsum}
\usepackage{makecell}

\usepackage{xcolor}

\DeclareMathOperator*{\argmax}{arg\,max}
\DeclareMathOperator*{\argmin}{arg\,min}

\newtheorem*{proof*}{Proof}
\newtheorem*{claim*}{}

\AtBeginDocument{%
  \providecommand\BibTeX{{%
    \normalfont B\kern-0.5em{\scshape i\kern-0.25em b}\kern-0.8em\TeX}}}

\newcolumntype{C}[1]{>{\centering\let\newline\\\arraybackslash\hspace{0pt}}m{#1}}
\newcommand\ChangeRT[1]{\noalign{\hrule height #1}}

\usepackage{listings}
\usepackage{color}
\definecolor{codegreen}{rgb}{0.3,0.5,0.0}
\begin{document}

\title{{Progressively} Optimized Bi-Granular Document Representation for {Scalable} Embedding Based Retrieval}

\author{
    Shitao Xiao$^{\textbf{{\tiny\ding{171}}}}$, Zheng Liu$^{\textbf{{\tiny\ding{168}}}}$, Weihao Han$^{\textbf{{\tiny\ding{169}}}}$, Jianjin Zhang$^{\textbf{{\tiny\ding{169}}}}$,
    Yingxia Shao$^{\textbf{{\tiny\ding{171}}}}$, Defu Lian$^{\textbf{{\tiny\ding{170}}}}$,  
    Chaozhuo Li$^{\textbf{{\tiny\ding{168}}}}$, Hao Sun$^{\textbf{{\tiny\ding{169}}}}$, Denvy Deng$^{\textbf{{\tiny\ding{169}}}}$, Liangjie Zhang$^{\textbf{{\tiny\ding{169}}}}$, Qi Zhang$^{\textbf{{\tiny\ding{169}}}}$, Xing Xie$^{\textbf{{\tiny\ding{168}}}}$\\                       
    {\ding{171}: Beijing University of Posts and Telecommunications,
    Beijing
    China} \\ 
    {\ding{168}: Microsoft Research Asia, Beijing, China}\\
    {\ding{169}: Microsoft STCA, Beijing, China}\\
    {\ding{170}: University of Science and Technology of China, Hefei, China}\\
}
\email{{zhengliu,weihan,jianjzh,cli,hasun,dedeng,liazha,qizhang,xingx}@microsoft.com}
\email{{stxiao,shaoyx}@bupt.edu.cn}
\email{liandefu@ustc.edu.cn}
    






\renewcommand{\shortauthors}{Xiao and Liu, et al.}
\renewcommand{\authors}{Shitao Xiao, Zheng Liu, Weihao Han, Jianjin Zhang, Yingxia Shao, Defu Lian, Chaozhuo Li, Hao Sun, Denvy Deng, Liangjie Zhang, Qi Zhang, Xing Xie}

\begin{abstract}
Ad-hoc search calls for the selection of appropriate answers from a massive-scale corpus. Nowadays, the embedding-based retrieval (EBR) becomes a promising solution, where deep learning based document representation and ANN search techniques are allied to handle this task. However, a major challenge is that the ANN index can be too large to fit into memory, given the considerable size of answer corpus. In this work, we tackle this problem with \textit{Bi-Granular Document Representation}, where the lightweight sparse embeddings are indexed and standby in memory for coarse-grained candidate search, and the heavyweight dense embeddings are hosted in disk for fine-grained post verification. For the best of retrieval accuracy, a \textit{Progressive Optimization} {framework} is designed. The sparse embeddings are learned ahead for high-quality search of candidates. Conditioned on the candidate distribution induced by the sparse embeddings, the dense embeddings are continuously learned to optimize the discrimination of ground-truth from the shortlisted candidates. Besides, two techniques: the {contrastive quantization} and the locality-centric sampling are introduced for the learning of sparse and dense embeddings, which substantially contribute to their performances. Thanks to the above features, our method effectively handles massive-scale EBR with strong advantages in accuracy: with up to $+4.3\%$ recall gain on million-scale corpus, and up to $+17.5\%$ recall gain on billion-scale corpus.
Besides, Our method is applied to a major sponsored search platform with substantial gains on revenue ($+1.95\%$), Recall ($+1.01\%$) and CTR ($+0.49\%$). Our code is available at https://github.com/microsoft/BiDR.
\end{abstract}


\maketitle

\section{Introduction}
Ad-hoc search is an important component for today's online applications, e.g., recommenders and advertising platforms. In response to each input query, the system needs to select appropriate answers from the entire corpus which may serve user's search demand. Nowadays, the embedding-based retrieval (EBR) has become a promising solution, where deep learning based document representation and ANN search techniques are jointly used to handle this problem. EBR consists of the following steps. In the offline stage, the text encoder is learned to represent the input documents as embeddings (e.g., DPR \cite{karpukhin2020dense}, ANCE \cite{xiong2021approximate}), where queries and their correlated answers can be projected close to each other in the latent space. Then, the ANN index (e.g., HNSW \cite{malkov2018efficient}, IVFADC \cite{jegou2011searching}) is built for the embeddings of the entire corpus. In the online stage, the ANN index is hosted in memory. Once a search request is presented, the input query is encoded into its embedding, with which relevant answers are efficiently selected by searching the ANN index. 

\begin{figure}[t]
\centering
\includegraphics[width=0.75\linewidth]{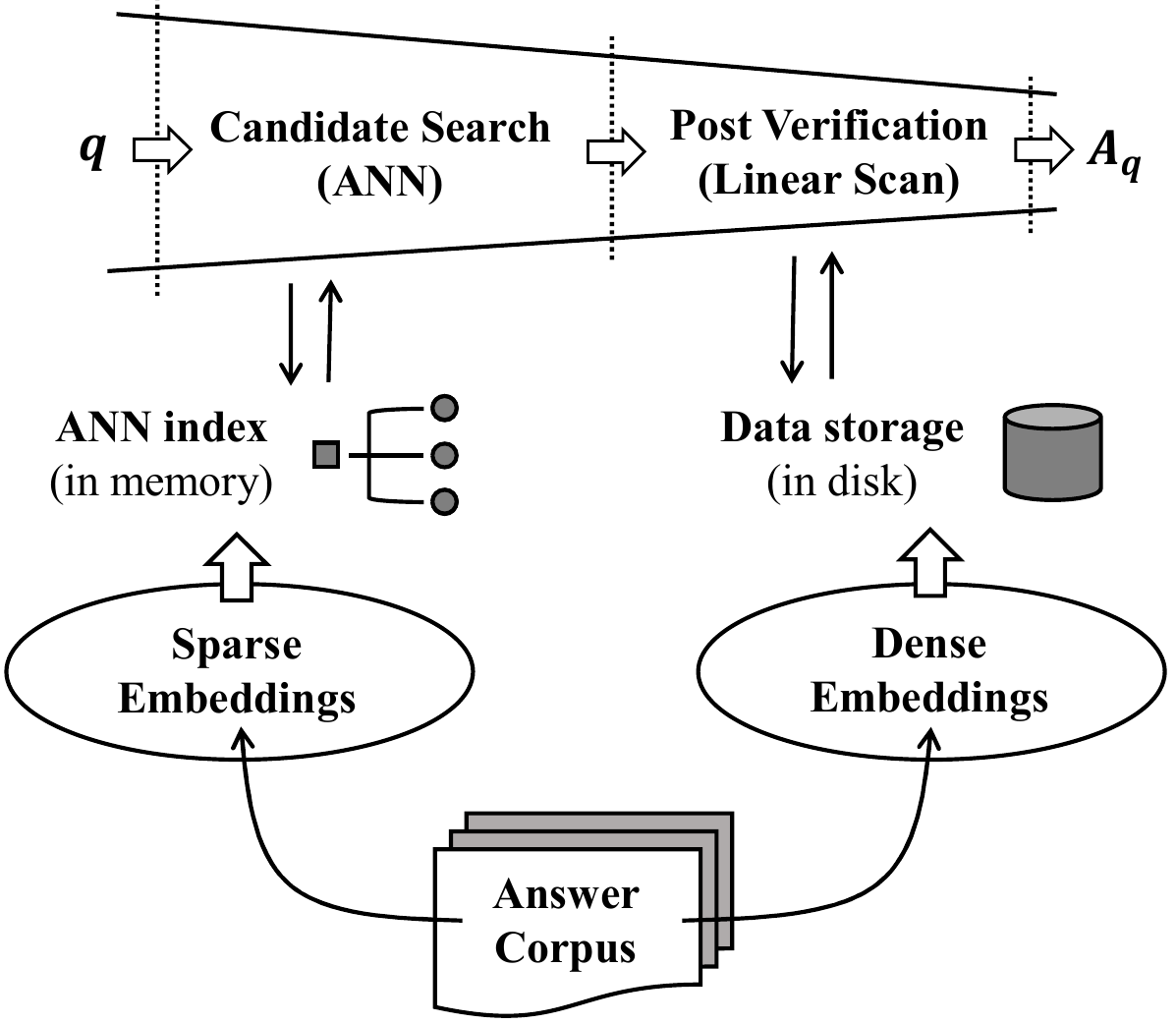}
\caption{\small Bi-granular document representation based EBR. The answers are represented as bi-granular embeddings: the sparse embeddings are indexed and standby in memory for candidate search; the dense embeddings are hosted in disk and used for post verification.}
\label{fig:1}
\end{figure}

\subsection{Embedding Based Retrieval At Scale}
EBR is challenging when a massive-scale corpus is given, as the ANN index can be too large to fit into memory; e.g., one billion dim-768 floating-point embeddings will take 3TB RAM space, which is orders of magnitude larger than the capability of a physical machine. Such a problem becomes even severe for today's online platforms, where billions of answers are generated by the content providers. One solution of scaling up EBR is to partition the data into multiple shards, each of which is hosted on a different machine. The query will be routed to all the shards, and the search results will be aggregated in the final stage \cite{fu2017fast}. However, the above method operates at a huge cost. In this work, we tackle this problem with the \textit{Bi-Granular Document Representation} (Figure \ref{fig:1}), where answers are represented as two sets of embeddings: the lightweight \textit{sparse embeddings}, which are indexed and standby in memory for coarse-grained candidate search; the heavyweight \textit{dense embeddings}, which are maintained in disk for fine-grained post verification. For each query $q$, the candidate answers are selected via in-memory ANN search; then, the dense embeddings of the shortlisted candidates are loaded from disk, with which the fine-grained post verification result $A_q$ is generated. With the above treatments, the EBR 
becomes scalable, where billions of embeddings can be hosted at merely tens of gigabytes RAM usage.

\begin{figure}[t]
\centering
\includegraphics[width=0.98\linewidth]{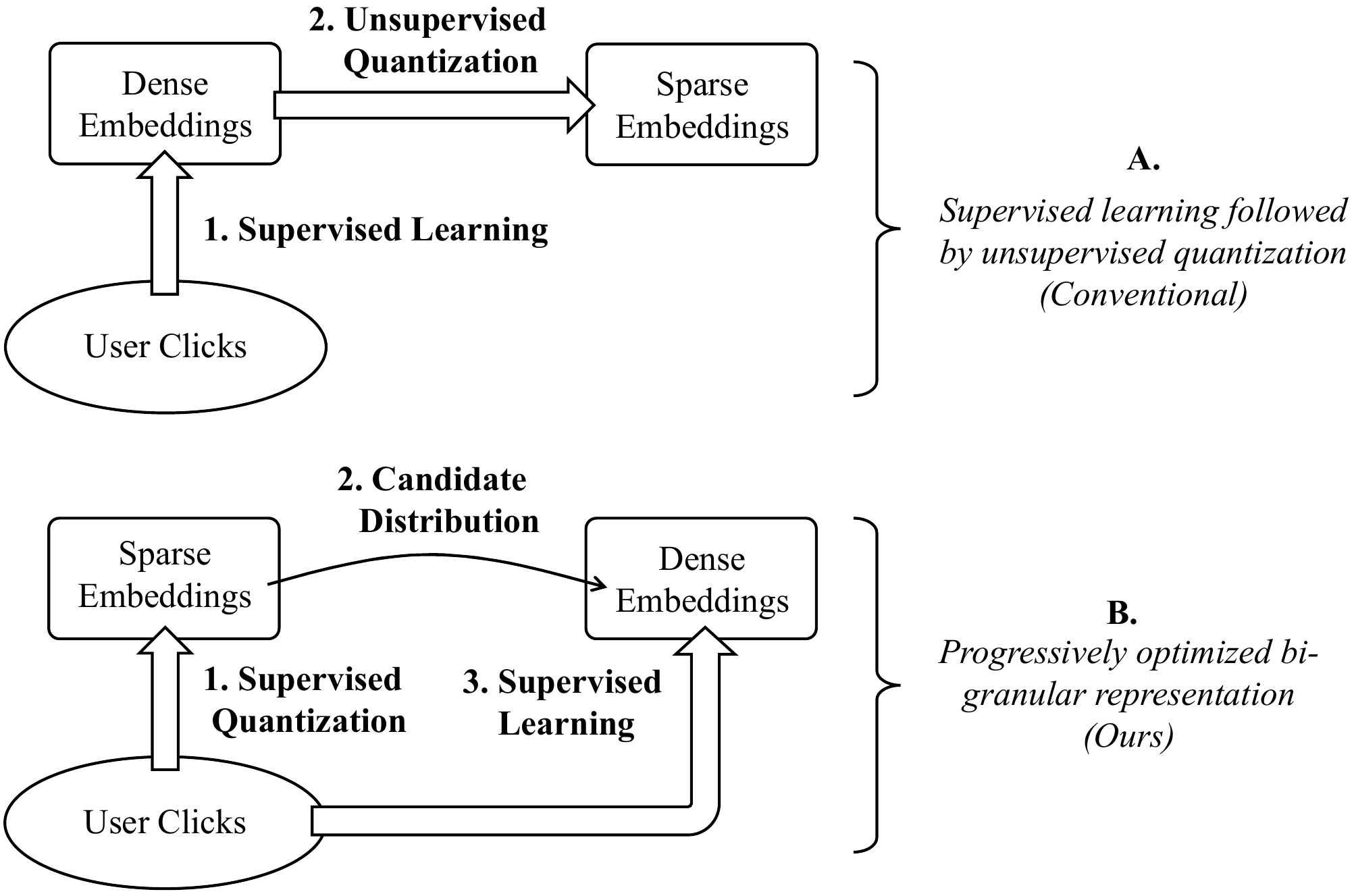}
\vspace{-2pt}
\caption{\small Comparison of bi-granular document representations. (A) The conventional method learns dense embeddings first and run unsupervised quantization for the sparse embeddings. (B) Our method learns the sparse embeddings first via supervision, based on which the dense embeddings are progressively optimized.}
\label{fig:1-0}
\end{figure}

\subsubsection{Limitations on representation learning}
Despite the satisfaction of memory space constraint, the learning of bi-granular document representation remains a challenging problem. The existing EBR algorithms typically learn the dense embeddings directly; when sparse embeddings are needed, they are unsupervisedly quantized from the well learned dense embeddings \cite{fan2019mobius,huang2020embedding,liu2021pre} (Figure \ref{fig:1-0} A). {However, we argue that the conventional way of generating bi-granular document representation is \textit{improperly optimized}, making it \textit{incompatible} with our massive-scale EBR workflow where the candidate-search and post-verification are successively performed.}


Particularly, the conventional dense embeddings are learned for the \textit{Global Discrimination}: the capability of finding a query's relevant answers from the entire corpus. However, considering that the dense embeddings are used for post verification, it needs to emphasize the \textit{Local Discrimination}: the capability of selecting the ground-truth from the shortlisted candidates. Therefore, it should be optimized on top of the candidate distribution induced by the sparse embeddings, which is different from the answer distribution over the entire corpus. Besides, it is also worth noting that the unsupervised compression is lossy {\cite{chen2020differentiable,xiao2021match}}, which prevents the full coverage of relevant answers in candidate search.

\subsection{{Progressively Optimized Representation}}
To address the above problem, we propose the \textit{Progressively Optimized Document Representation}, with learning objectives corrected for the {``candidate-search plus post-verification'' based retrieval} ({Figure \ref{fig:1-0} B}). The sparse embeddings are supervisedly learned first: instead of running unsupervised quantization for the well-learned dense embeddings, the sparse embeddings are generated from \textit{contrastive learning}, which optimizes the \textit{global discrimination} and helps to enable high-quality answers to be covered in candidate search. More importantly, the sparse embeddings will formulate the candidate distribution, on top of which the dense embeddings can be optimized for the \textit{local discrimination}. In this place, a novel sampling strategy called \textit{locality-centric sampling} is introduced to facilitate the effective optimization of local discrimination: a bipartite proximity graph is established first for the queries and answers based on the distance measured by sparse embeddings; then, while generating the training instances, a random query is selected as the entry, starting from which queries and answers are sampled over a local {area} on graph. In this way, the negative samples within the same batch become locally concentrated. By learning to discriminate the ground-truth from the massive amount of ``in-batch shared negative samples'', the representation model will be equipped with superior capability on local discrimination, which substantially contributes to the post verification's accuracy.

\subsubsection{Highlighted Merits} 
Thanks to the above features, the retrieval accuracy can be notably improved for massive-scale EBR. Besides, we surprisingly find that our method even outperforms the SOTA EBR methods in direct comparisons (i.e., the SOTA baselines directly use dense embeddings for retrieval tasks, where no quantization loss is incurred). Knowing that the sparse embeddings can be served with extremely small costs, such an advantage indicates that our bi-granular document representation may also be applied to generic EBR, where a moderate-scale corpus is given. 


Comprehensive experimental studies are performed with popular benchmarks on web search and a billion-scale dataset on sponsored search. According to the experimental results, our method achieves notable improvements against the SOTA baselines in both massive-scale and generic EBR scenarios. Meanwhile, it fully maintains the high scalability and running efficiency. Our method is also applied to a commercial search engine, which brings strong revenue and CTR gains to its advertising service. Finally, The contributions of this paper are highlighted as follows.
\begin{itemize}
    \item We tackle the massive-scale EBR problem on top of 
    the progressively optimized bi-granular document representation. To the best of our knowledge, this is the first systematic work of its kind, which notably improves the retrieval accuracy.
    \item Leveraging the contrastive quantization and locality-centric sampling, the global and local discrimination can be effectively optimized for the sparse and dense embeddings.
    \item The comprehensive experimental studies verify the effectiveness of our proposed method in both massive-scale and generic EBR scenarios.
\end{itemize}

We'll present the related work, methodology, and experimental studies in the rest part. The supplementary material is also provided, showing more details like notations and additional experiments.

\section{Related Work}
In this section, we review two fundamental techniques of embedding-based retrieval: the deep learning based document representation, and the indexing algorithms for scalable vector search.


$\bullet$ \textbf{Deep Document Representation}. Document representation is a fundamental part in EBR. Recently, deep learning based methods thrive thanks to the generation of semantic-rich embeddings \cite{shen2014learning,le2014distributed}. The latest methods intensively explore the pretrained language models, e.g., BERT \cite{devlin2018bert}, RoBERTa \cite{liu2019roberta}, XLnet \cite{yang2019xlnet}. Such models leverage large-scale transformers as backbones and get fully pretrained with massive corpus, which achieves strong expressiveness, and contributes to numerous downstream tasks, such as document retrieval \cite{xiong2021approximate,karpukhin2020dense,chang2020pre} and question answering \cite{lee2019latent,joshi2020spanbert,guu2020realm}.

Besides, substantial progress has also been made in training methods. One such area is negative sampling. It is found that the representation learning is influenced by the scale of negative samples. Thus, the in-batch negative sampling is introduced \cite{chen2020simple,karpukhin2020dense,gao2021simcse}, where negative samples can be augmented in a cost free way. Later on, the cross-device negative sampling is utilized in \cite{qu2021rocketqa,xiao2021match}, where negative samples can be shared across all the distributed devices. Recent studies also prove that the usefulness of ``hard'' negative samples. In \cite{luan2021sparse,karpukhin2020dense}, hard negatives are collected by BM25 score. In \cite{xiong2021approximate,zhan2021optimizing,qu2021rocketqa}, even harder negatives are collected from ANN search. With the increased hardness, the representation quality is reportedly improved by notable margins. Finally, efforts have also been paid to positive samples. In \cite{chang2020pre,gao2021simcse,zhang2020unsupervised}, positives are introduced by the self-supervision algorithms. Following the idea of knowledge distillation, the representation models are learned w.r.t. a more expressive teacher \cite{lu2020twinbert,lin2020distilling,qu2021rocketqa} (usually a cross encoder): with pseudo labels scored for the given candidate list, additional supervision signals can be provided for the representation model (\textit{i.e.}, the student).

$\bullet$ \textbf{Indexing Algorithms}. A major building block of EBR is the ANN index. In short, the ANN index organizes high-dimensional vectors with certain data structures such that the proximity search can be performed with sub-linear time complexities. The ANN index has been studied for a long time. Many of the early works were developed based on space partitioning; e.g., LSH \cite{datar2004locality}, BBF KD-Tree \cite{beis1997shape}, K-means Tree \cite{muja2009fast}. Recently, the quantization-based approaches received extensive attention thanks to the effectiveness and compactness; e.g., in IVFADC \cite{jegou2010product,jegou2011searching}, the inverted index is combined with PQ (product quantization) to facilitate fast retrieval of ANN query. Another important class of ANN index is the graph-based techniques, where the proximity graph is built for the entire corpus. One such method is HNSW \cite{malkov2018efficient}: a multi-layer structure of hierarchical proximity graphs are constructed, which achieve effective retrieval of the nearest neighbours with logarithmic complexity. Later on, several works have been proposed for the better construction of proximity graphs, e.g., NSG \cite{fu2017fast} and Vamana \cite{subramanya2019diskann}, which lead to improved search performance. The graph-based technique is also combined with quantization-based approach, e.g., IVFOADC+G+P \cite{baranchuk2018revisiting}, where the proximity graph is constructed for the fast retrieval of top-$K$ centroids from a large inverted index. As of this writing, the graph-based methods remain the most competitive ANN algorithms in terms of recall and efficiency.


One challenging problem of ANN is that the index can be oversized for the memory, when facing a massive-scale corpus. Although quantization based methods can help, they are reported to be limited in recall \cite{subramanya2019diskann,ren2020hm}. The latest approaches resort to hybrid storage. In DiskANN \cite{subramanya2019diskann,singh2021freshdiskann}, the Vamana graph is built for the quantized embeddings, and is maintained in RAM for cost-grained search; meanwhile, the original embeddings are kept in disk for post verification. In HM-ANN \cite{ren2020hm}, the NSW graph is maintained in slow memory (PMM) and the pivot points are hosted in fast memory (DRAM). However, the requirement of massive slow memory in HM-ANN is expensive and unavailable in many situations. So far, DiskANN remains the general solution for large-scale ANN due to its relative competitiveness in precision and serving cost.

\begin{figure}[t]
\centering
\includegraphics[width=0.99\linewidth]{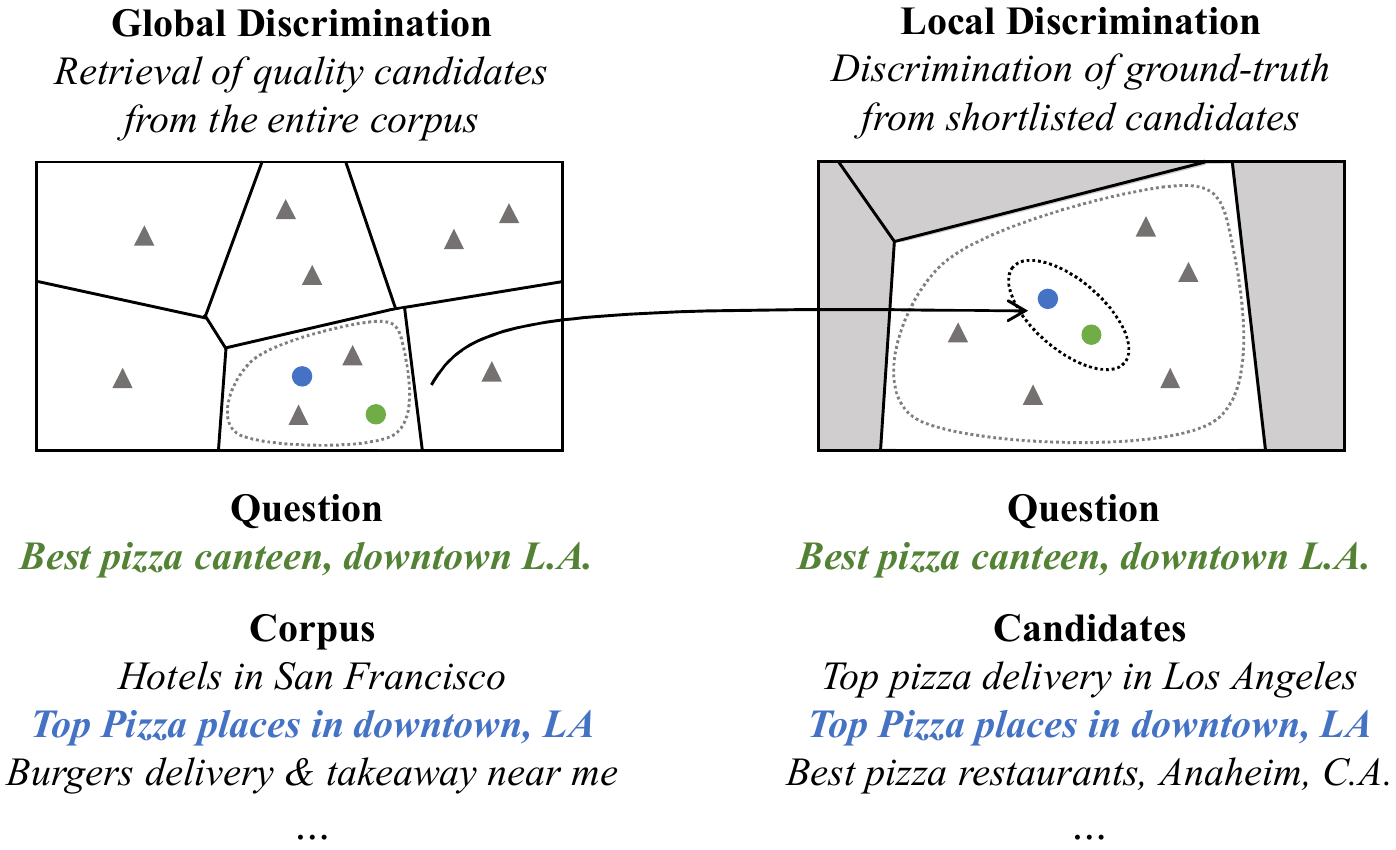}
\vspace{-10pt}
\caption{\small Progressive optimization process. Left: the sparse embeddings are learned first to optimize the global discrimination, which makes high-quality candidates to be retrieved. Right: the dense embeddings are learned to optimize the local discrimination conditioned on the candidate distribution, which ensures the effective discrimination of ground-truth (blue).}
\label{fig:3-0}
\end{figure}

\vspace{-5pt}
\section{Methodology}
{The overall learning process is shown as Figure \ref{fig:3-0}. In the first stage, the sparse embeddings are learned to 
optimize the \textit{global discrimination}: to distinguish the right answers from the irrelevant ones in the entire corpus, such that query (green) and its related answers can be confined in the same laten area. In the second stage, the dense embeddings are learned to optimize the \textit{local discrimination}: to identify the ground-truth from the shortlisted candidates promoted by the sparse embeddings.}



\vspace{-5pt}
\subsection{Sparse Embedding Learning}
{We will first present the encoding network for the sparse embeddings. Then, we'll formulate the contrastive learning to optimize the global discrimination, together with its negative sampling strategy.}

\subsubsection{ADC Siamese {Network}} 
{Note that the memory-efficient representation is only needed for the answers, which call for indexing and in-memory storage. Thus, the answers are quantized for sparse representation, while the queries remain densely represented}. We adapt the siamese {network} with asymmetric distance computation (ADC) \cite{jegou2011searching}, in which the sparse embedding relies on its selected codewords to compute the distance with the dense embedding.
Particularly, the answer's encoding involves the following steps. Firstly, each answer $a$ is projected into its dense representation. Following the recent works on document representation \cite{karpukhin2020dense,luan2021sparse}, we use BERT as the encoding backbone. The generated dense vector is denoted as $\mathbf{z}^a$. Secondly, the dense vector is quantized into binary codes. Here, the quantization is made through a differentiable PQ component \cite{chen2020differentiable}. The quantizer is parameterized with $M$ codebooks, each codeword contains $P$ codewords: $\{\mathbf{C}_{i,j}\}_{M,P}$. Given the dense representation $\mathbf{z}^a$, a codeword is selected from each of the codebooks:
\begin{equation}\label{eq:4}
    \mathbf{C}^*_i = \argmax_{j} \{ \langle \mathbf{z}^a_i, \mathbf{C}_{i,j} \rangle \},
\end{equation}
where $\mathbf{z}^a_i$ is the $i$-th segment of $\mathbf{z}^a$. The argmax operator is realized based on the straight through estimation \cite{bengio2013estimating}, such that it is end-to-end differentiable. The sparse embedding for the answer consists of $M$ one-hot vectors, corresponding to the ID of each selected codeword. Finally, the sparse embedding concatenates all the selected codewords for distance computation:
\begin{equation}\label{eq:5}
    f_s(a) = \mathrm{concat}([\mathbf{C}^*_1,...,\mathbf{C}^*_M]).
\end{equation}
The query is directly encoded by BERT, which is the same as the intermediate result $\mathbf{z}^a$ in answer encoding.

\subsubsection{Contrastive learning} Instead of running unsupervised quantization on the dense embeddings, the sparse embeddings are supervisedly learned for the optimized searching accuracy. As discussed, the sparse embeddings are used to select quality candidates from the entire corpus, which is illustrated by the following operation:
\begin{equation}\label{eq:1}
    \widehat{{A}}_q \leftarrow \text{Top-N}(\{\langle g(q), f_s(a) \rangle|A\}).
\end{equation}
In this place, $\widehat{{A}}_q$ is the candidate set, whose size N is much smaller than the corpus scale, \textit{i.e.}, $\text{N} \ll |A|$. $g(\cdot)$ is the query's encoder and $f_s(\cdot)$ is the sparse answer encoder. We use inner product $\langle\cdot\rangle$ for the measurement of embedding similarity. The supervised learning is performed to minimize the following loss:
\begin{equation}\label{eq:2}
    \argmin\limits_{g, ~f_s} \sum_q \sum_{a^+} \sum_{a^-} l(g(q), f_s(a^+), f_s(a^-)),
\end{equation}
where $a^+$ and $a^-$ are the positive and negative answers to the query, respectively. Here, we use InfoNCE \cite{oord2018representation} as our loss function:
\begin{equation}\label{eq:3}
    \sum_{a^-} l(q, a^+, a^-) \leftarrow -\log \frac{\exp(\langle g(q), f_s(a^+) \rangle)}{\sum_{a^+\cup A_q^-}\exp(\langle g(q), f_s(a) \rangle)},
\end{equation}
where $A_q^-$ denotes the set of negative samples w.r.t. $q$ and $a^+$. 

\subsubsection{Negative Sampling}\label{sec:negative_stage1}
Knowing that the scale of negative samples substantially contributes to the representation quality \cite{chen2020simple,qu2021rocketqa}, we leverage in-batch negatives \cite{karpukhin2020dense}, where the positive answer in one training instance is used for the augmentation of other instance's negative samples. Besides, we further sample one lexically similar negative for each query based on BM25, which is proved to be helpful for document representation \cite{luan2021sparse,karpukhin2020dense}. The lexically similar negatives are also shared across the training instances. With the above treatments, the negative sample set $A_q^-$ in Eq. \ref{eq:3} becomes:
\begin{equation}\label{eq:6}
    A_q^- = \{a^+\}_{q' \neq q} \cup \{a^-\}_{B},
\end{equation}
where $\{a^+\}_{q' \neq q}$ denotes the positive answers from other queries $q'$, and $\{a^-\}_{B}$ means all the BM25 negatives within the same mini-batch batch. That is to say, each query may get almost $2|B|$ negative samples ($|B|$ is the batch size). Notice that we may also collect even ``harder'' negative samples than the lexical-similar ones with ANN-based negative sampling \cite{xiong2021approximate,qu2021rocketqa,zhan2021optimizing}; however, {we experimentally find that having such treatments in the first stage does not contribute to the overall precision after post verification}, given that the local discrimination will be fully optimized for the dense embeddings. As a result, we stay with the BM25 based negative samples for its effectiveness and simplicity. 


\begin{figure}[t]
\centering
\includegraphics[width=0.75\linewidth]{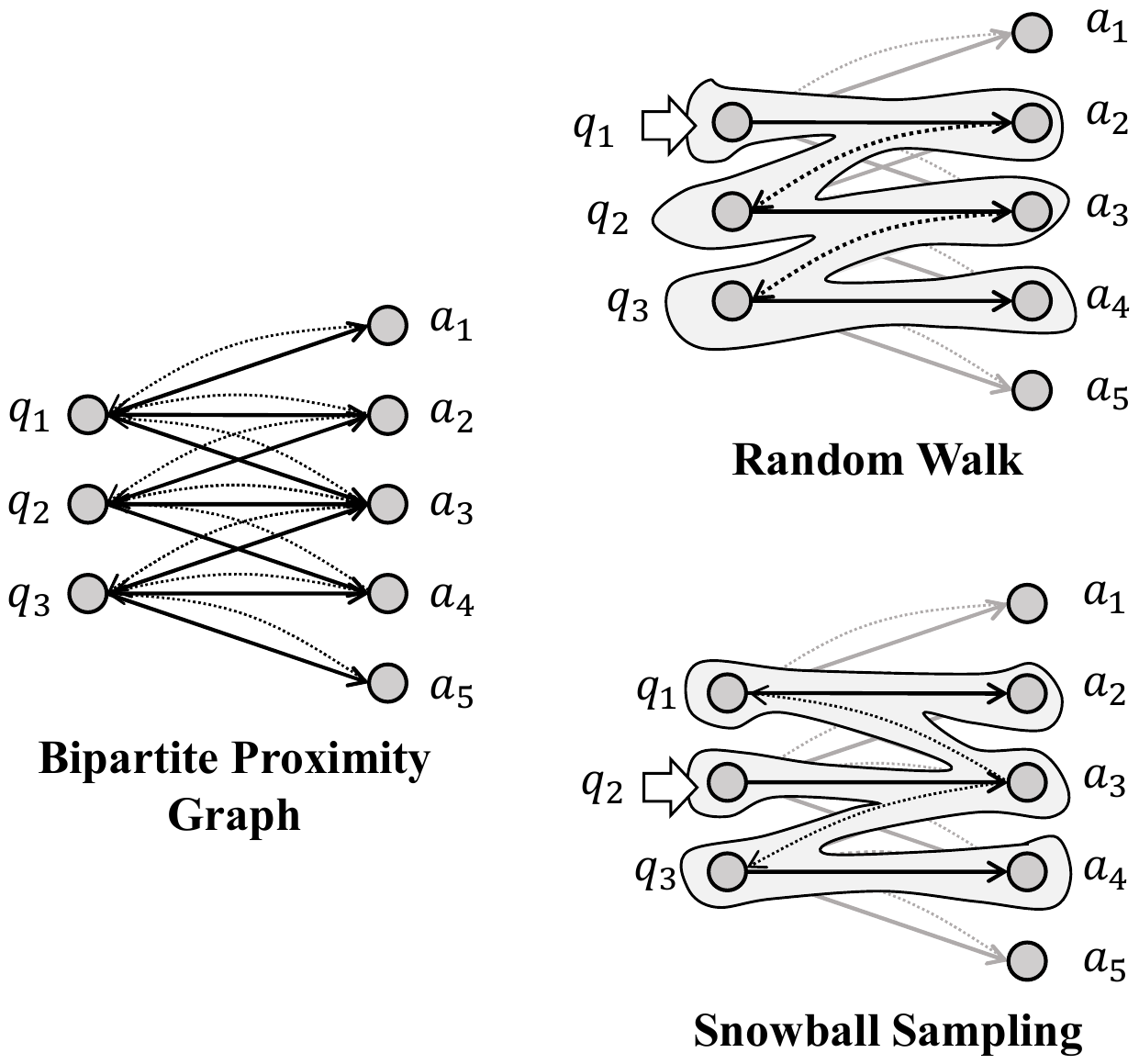}
\caption{\small {Locality Centric Sampling.} (1) bipartite proximity graph construction: each query connects to its top-N candidate answers; each answer connects back to its original query. (2) Starting from the entry query, the training instances are sampled on a local patch via two optional strategies: random walk and snowball sampling.}
\label{fig:3}
\end{figure}

\subsection{Dense Embedding Learning}
{We rely on the siamese encoder with BERT backbone to generate the dense embeddings. The supervision objective is adapted, and the locality-centric sampling (LCS) is introduced to realize strong local discrimination. Finally, a post processing treatment is introduced to unify the query encoders learned in the first and second stage.}

\subsubsection{Supervision objective}
The dense embeddings are learned conditionally w.r.t. the candidate distribution induced by the sparse embeddings: to identify the ground-truth from the candidates promoted by the sparse embeddings. Such a capability is called the local discrimination, where the InfoNCE loss is adapted as follows:
\begin{equation}\label{eq:7}
    \sum_{a^-} l(q, a^+, a^-) \leftarrow -\log \frac{\exp(\langle g'(q), f_d(a^+) \rangle)}{\sum_{a^+\cup \widehat{A}_q^-}\exp(\langle g'(q), f_d(a) \rangle)}.
\end{equation}
Here, $g'(\cdot)$ and $f_d(\cdot)$ are the query encoder and dense answer encoder (we'll learn a new query encoder $g'(\cdot)$ first, and then unify $g'(\cdot)$ and $g(\cdot)$ in the post processing). $\widehat{A}_q^-$ are negative samples collected for this stage, which are different from the ones used in first stage, as Eq. \ref{eq:6}. Particularly, $\widehat{A}_q^-$ become a subset of answers randomly sampled from the candidate set:
\begin{equation}
    \widehat{A}_q^- \leftarrow \mathcal{S}(\widehat{A}_q).
\end{equation}
Here, $\widehat{A}_q$ is the top-N candidates generated as Eq. \ref{eq:1}, and $\mathcal{S}(\cdot)$ is the random selection operator. We empirically find that $|\widehat{A}^-_q|$ should be large enough in order to effectively optimize the local discrimination.  However, one challenging problem of having a large $|\widehat{A}^-_q|$ is the heavy encoding cost. In many cases, it is infeasible to keep a large sample size as the GPU capacity is limited and the encoding cost per answer can be high. To overcome this problem, the locality-centric sampling is proposed to expand $\widehat{A}^-_q$ in a cost-efficient way. 


\begin{algorithm}[t]
\caption{Training}\label{alg:1}
    \LinesNumbered 
    \SetKwInOut{KwIn}{Input}
    \SetKwInOut{KwOut}{Output}
    \KwIn{$\{(q,a^+)\}$, $A$}
    \KwOut{$f_s$, $f_d$, $g$, $g'$, $g''$}
    \Begin{
        \% (\textit{Learn sparse embeddings})\\
        $f_s$, $g$ $\leftarrow$ argmin InfoNCE in Eq. \ref{eq:3}\; 
        \% (\textit{Learn dense embeddings})\\
        construct bipartite proximity graph $\mathcal{G}$\;
        \While{not converge}{
        get mini-batch $B$ by performing LCS on $\mathcal{G}$\;
        update $f_d$, $g'$ by minimizing InfoNCE in Eq. \ref{eq:7} on $B$\;
        }
        $g''$ $\leftarrow$ argmin InfoNCE in Eq. \ref{eq:9}\;
    }
\end{algorithm}

\subsubsection{Locality Centric Sampling}
We introduce LCS for the augmentation of $\widehat{A}^-_q$. Firstly, the Bipartite Proximity Graph is constructed, which organizes the queries and answers based on their sparse embedding similarity. Secondly, the random sampling is performed on a local patch of the graph, which iteratively generates queries and their corresponding hard negatives. By doing so, the negative sample of one query may also be shared as a hard negative sample to another query within the same mini-batch. Details about LCS are introduced as follows.

$\bullet$ \textbf{Bipartite Proximity Graph}. We first construct the bipartite proximity graph for the queries and answers based on their sparse embeddings similarity, as Figure \ref{fig:3}. Particularly, the Top-N candidate answers $\widehat{A}_q$ are selected for each query, as Eq. \ref{eq:1} ({N=200 while constructing the graph}). Then, a forward edge is added (solid line), which connects query $q$ to an answer $a$ in $\widehat{A}_q$. Besides, a backward edge is also added (dash line), which points back to $q$ from $a$.

$\bullet$ \textbf{Sampling on Graph}. Starting from a random query $q$ as the entry point, one of the following strategies: {Random Walk} and {Snowball Sampling} (Figure \ref{fig:3}), is performed to generate a mini-batch of training instances. Each strategy has its pros and cons in different situations, which will be discussed in our experiments.

\textbf{Random Walk}: the sampling follows a zigzag path. For each visited query $q_i$, a random answer $a_i^-$ is selected from $q_i$'s connected candidates (with ground-truth excluded); then, another query $q_{i+1}$ is sampled from the backward connections of $a_i^-$. It can be comprehended that the next answer $a_{i+1}^-$ sampled for query $q_{i+1}$ is similar with $a_i^-$; therefore, it is a potential candidate to $q_i$, which will contribute to the local discrimination. 

\textbf{Snowball Sampling}: {one visited query $q$ will sample a random answer $a^-$ from its neighbours; then all directly connected queries of $a$ will be visited in the next step.} Compared with random walk, the training instances collected by snowball sampling can be more concentrated within a local area, which improves the in-batch samples' hardness but potentially reduces the diversity.


The sampling process is consecutively run to generate $\{q, a^+, a^-\}$ (the positive answer $a^+$ is directly collected when $q$ is visited). The sampling completes when the mini-batch is filled up by the collected instances. Finally, the negative answers for a query $q$ will be: $\{a^+\}_{q' \neq q}$ (others' positives) plus $\{a^-\}_B$ (negatives within the batch), which means each query will get as many as $2|B|$-1 locally concentrated negative samples.


\subsubsection{Query Unification}
So far, we've learned two query encoders $g(\cdot)$ and $g'(\cdot)$ paired with the sparse and dense answer embeddings, respectively. Although this may not cause extra latency in online serving, as both encoders may run in parallel, it still incurs twice encoding cost. In case the computation capacity is limited, we unify the query encoders in a simple but effective way, so that each query only needs to be encoded once. Particularly, we initialize $g''(\cdot)$: $g''(\cdot) \leftarrow g'(\cdot)$, and fix the answer encoders: $f_s(\cdot).fix$, $f_d(\cdot).fix$ (the fixed answers let us maintain a massive negative sample queue \cite{he2020momentum}, which benefits the fine-tuning effect). Then, we finetune the query encoder to minimize the combined InfoNCE loss:
\begin{equation}\label{eq:9}
\begin{split}
    \argmin\limits_{g''} & \sum\nolimits_q\sum\nolimits_{a^+} l(g''(q),f_s(a^+).fix, f_s(a^-).fix|A^-_q) + \\
    & \sum\nolimits_q\sum\nolimits_{a^+} l(g''(q),f_d(a^+).fix, f_d(a^-).fix|\widehat{A}^-_q).
\end{split}
\end{equation}
The upper part is the InfoNCE loss in Eq. \ref{eq:3}; the lower part is the InfoNCE loss in Eq. \ref{eq:7}. By optimizing the above combined loss, the unified query encoder $g''(\cdot)$ is learned. We find that the original precision can be well-preserved after the unification. 

\begin{algorithm}[t]
\caption{ANN Search}\label{alg:2}
    \LinesNumbered 
    \% (\textit{Offline Index Construction}) \\
    \Begin{
        infer sparse \& dense embeddings: $\{f_s(a)|A\}$, $\{f_d(a)|A\}$\;
        build HNSW with $\{f_s(a)|A\}$ and host it in memory\;
        maintain $\{f_d(a)|A\}$ in disk\;
    }
    \% (\textit{Online ANN Serving}) \\
    \Begin{
        encoder query as $g''(q)$\;
        $\widehat{A}_q$ $\leftarrow$ search HNSW for Top-N($\{\langle f_s(a), g''(q) \rangle| A\}$)\;
        load $\{f_d(a)|\widehat{A}_q\}$ into memory\;
        $A_q$ $\leftarrow$ post re-rank for Top-K($\{\langle f_d(a), g''(q) \rangle| \widehat{A}_q\}$)\;
    }
\end{algorithm}

\subsection{Training and Search Algorithm}
The training is summarized as Alg. \ref{alg:1}. Firstly, the sparse answer encoder $f_s(\cdot)$ and the query encoder $g(\cdot)$ are learned to minimize the InfoNCE loss in Eq. \ref{eq:3}. Secondly, the bipartite proximity graph $\mathcal{G}$ is constructed. The mini-batch $B$ is sampled by performing LCS on $\mathcal{G}$, with which the dense encoder $f_d(\cdot)$ and the query encoder $g'(\cdot)$ are learned w.r.t. the InfoNCE loss in Eq. \ref{eq:7}. Finally, the
query encoder is unified as $g''(\cdot)$ w.r.t. the joint InfoNCE in Eq. \ref{eq:9}.

The ANN search (Alg. \ref{alg:2}) involves two parts: the offline index construction, and the online serving of ANN request. For the offline stage: the sparse embeddings $f_s(a)$ and dense embeddings $f_d(a)$ are inferred for the entire answers $A$. Then, HNSW \cite{malkov2018efficient} is constructed for $f_s(a)$, which is standby in memory. Meanwhile, all the dense embeddings $\{f_d(a)|A\}$ are stored in disk. For online ANN serving: each input query $q$ is encoded as $g''(q)$, which is used to search HNSW for the Top-N candidates $\widehat{A}_q$. Then, the dense embeddings are loaded into memory for the candidates. Finally, the fine-grained Top-K result $A_q$ is generated by ranking $\hat{A}_q$ based on $\{f_d(a)|\widehat{A}_q\}$. Note that when computation capacity is sufficient, we'll still maintain two query encoders $g(q)$ and $g'(q)$ running in parallel, which may slightly benefit the retrieval accuracy.


\begin{table*}[t]
    \centering
    \footnotesize
    \begin{tabular}{p{2.2cm} | C{1.2cm} C{1.2cm} C{1.2cm} | C{1.2cm} C{1.2cm} C{1.2cm} | C{1.2cm} C{1.2cm} C{1.2cm}  }
    \ChangeRT{1pt}
    & \multicolumn{3}{c}{\textbf{Web Search (P)}} & \multicolumn{3}{|c}{\textbf{Web Search (D)}} & \multicolumn{3}{|c}{\textbf{Sponsored Search}} \\
    \cmidrule(lr){1-1}
    \cmidrule(lr){2-4}
    \cmidrule(lr){5-7}
    \cmidrule(lr){8-10}
    \textbf{Methods} & 
    \textbf{R@10} & \textbf{R@50} & \textbf{R@100} & \textbf{R@10} & \textbf{R@50} & \textbf{R@100} & \textbf{R@10} & \textbf{R@50} & \textbf{R@100}  \\
    \hline
    DPR + DISK & 0.5297 & 0.7610 & 0.8317 & 0.5969 & 0.8083 & 0.8623 & 0.3205 & 0.5043 & 0.5545  \\
    STAR + DISK & \underline{0.5834} & \underline{0.7807} & \underline{0.8331} & \underline{0.6807} & \underline{0.8575} & \underline{0.8925} & \underline{0.3438} & \underline{0.5480} & \underline{0.6007}  \\
    ANCE + DISK & 0.5745 & 0.7804 & \underline{0.8331} & 0.6680 & 0.8336 & 0.8659 & 0.3258 & 0.5247 & 0.5757  \\
    DE-BERT + DISK & 0.5537 & 0.7682 & 0.8265 & 0.5998 & 0.8089 & 0.8563 & 0.3279 & 0.5174 & 0.5659  \\
    \hline
    \textbf{BiDR (RD)} & 0.6048 & 0.8156 & 0.8753 & \textbf{0.6988} & \textbf{0.8821} & 0.9176 & 0.3985 & 0.6467 & 0.7152  \\
    \textbf{BiDR (SN)} & \textbf{0.6087} & \textbf{0.8169} & \textbf{0.8786} & 0.6959 & 0.8808 & \textbf{0.9225} & \textbf{0.4038} & \textbf{0.6489} & \textbf{0.7158}  \\
    \ChangeRT{1pt}
    \end{tabular}
    \caption{\small Evaluation of massive-scale EBR. BiDR (RD) and (SN) outperform all baselines with notable advantages.}
    \vspace{-22pt}
    \label{tab:1}
\end{table*}

\section{Experimental Studies}
The experimental studies focus on five issues. First and foremost, the evaluation of massive-scale EBR. Secondly, the evaluation of generic EBR (with a moderate-scale corpus). Thirdly, the efficiency and scalability. Fourthly, the impact from each of the proposed techniques. Finally, the impact to real-world search platforms. Part of the settings and additional results are shown in the appendix due to limited space. 


\subsection{Settings}
\subsubsection{Datasets} The TREC 2019 Deep Learning (DL) Track is a widely used benchmark on web search, which is taken as our primary experimental subject. The benchmark includes two tasks: the {passage} retrieval \textbf{Web Search (P)}, with a corpus of 8,841,823 answers; and the {document} retrieval \textbf{Web Search (D)}, with a corpus of 3,213,835 answers. We follow the official settings in \cite{craswell2020overview}. Besides, we utilize an industrial dataset on \textbf{Sponsor Search}, whose corpus contains one billion ads from a commercial search engine. As far as we know, this might be the largest corpus used in relevant studies, which helps us better analyze the performance in massive-scale EBR. In this dataset, each query needs to match the ground-truth ad (clicked by the user). There are 3,000,000 queries for the training set and 10,000/10,000 queries for the valid/test sets. Each query has one ground-truth advertisement. Note that the datasets are highly diversified in answers' lengths: Web Search (D) has an average length of 448.5 tokens, which is much longer than Web Search (P) (79.6) and Sponsored Search (8.1). Such a difference brings substantial impact to some of the experiment, which will be shown in our analysis.

\vspace{-5pt}
\subsubsection{Baseline Methods}
{We make comparison with the latest document representation methods, which achieve strong performances on tasks like web search and question answering.} 1) \textbf{DPR} \cite{karpukhin2020dense}, where in-batch negative samples are used for the training of siamere BERT based document encoders. It achieves competitive dense retrieval performance on open domain QA tasks. 2) \textbf{DE-BERT} \cite{luan2021sparse}, which adapts DPR by introducing extra hard negative samples. 3) \textbf{ANCE} \cite{xiong2021approximate}, which iteratively mines hard negatives globally from the entire corpus. It is one of the strongest dense retrieval baselines on MS MARCO benchmark. 4) \textbf{STAR} \cite{zhan2021optimizing}, which improves ANCE by introducing in-batch negatives for stabilized and more accurate document representation. We use the following ANN indexes in the experiments. To evaluate the massive-scale EBR performance, the SOTA scalable ANN algorithm \textbf{DiskANN} \cite{subramanya2019diskann} is leveraged. To evaluate the generic EBR performance, we adopt \textbf{HNSW} \cite{malkov2018efficient}, which is one of the most effective ANN algorithms in practice.  

\vspace{-5pt}
\subsection{Analysis}
\label{sec:analysis}
\subsubsection{Evaluation of massive-scale EBR}\label{sec:exp-1}
The evaluations of massive-scale EBR are shown in Table \ref{tab:1}, in which the baseline document representations are combined with DiskANN (\textbf{$\#\#$ + DISK}): the dense embeddings are compressed by PQ and indexed by Vamana graph. Our method (\textbf{BiDR}: \textbf{Bi}-granular \textbf{D}ocument \textbf{R}epresentation) has two forms: \textbf{BiDR (RD)}, where the LCS is performed via Random Walk; and \textbf{BiDR (SN)}, where the LCS is performed via Snowball sampling. Both baselines and our methods receive 1000 answers from candidate search, and re-rank them for the fine-grained Top-$K$ answers. According to the experiment results, both BiDR (RD) and (SN) outperform the baselines with consistent and notable advantages; e.g., the \textbf{Recall@10} can be relatively improved by as much as \textbf{4.34}\% and \textbf{2.66}\% on Web Search (P) and (D), compared with the strongest baselines. Given that Sponsored Search has a billion-scale corpus (more than 100$\times$ larger than Web Search), BiDR's advantages in recall rate become expanded, e.g., the \textbf{Recall@10} can be relatively improved by \textbf{17.45}\%.

More detailed analysis is made as follows. The performance of BiDR (SN) is better than BiDR (RD) on Web Search (P) and Sponsored Search, but becomes worse than BiDR (RD) on Web Search (D). One notable difference is that Web Search (P) and Sponsored Search have much smaller answer lengths (79.6 and 8.1) than Web Search (D) (448.5). Thus, the encoding of each training instance (one query with its positive and negative answers) takes much smaller memory consumption for both datasets. Knowing that the batch size is up-bounded by the GPU RAM capacity, it will call for a large amount of LCS sampling steps to fill up a mini-batch when applied to Web Search (P) and Sponsored Search. As a result, the random walk may get divergent and introduce a great deal of ``remote negative samples'', which is unfavorable to the local discrimination. In contrast, snowball helps to keep the sampling concentrated, which alleviates this problem. However, the number of LCS steps becomes highly limited in Web Search (D), given the large memory consumption per training instance. In this case, the random walk helps to introduce more ``diversified hard negative samples'', which provides more informative supervision signals than the snowball sampling.

As for baselines: DE-BERT is stronger than DPR; meanwhile, ANCE further outperforms DE-BERT. It indicates hard negatives, especially the ANN hard negatives, are beneficial. Besides, STAR achieves the highest baseline performance, where in-batch and ANN negatives are jointly used. {Our experiments in \ref{sec:exp-4} will show that the effects of ANN and in-batch negatives are different: the former one is more related with the local discrimination, while the latter one is more related with the global discrimination. The two objectives may become contradicted in certain situations.} 


\begin{table*}[t]
    \centering
    \small
    \footnotesize
    \begin{tabular}{p{2.2cm} | C{1.1cm} C{1.1cm} C{1.1cm} C{1.1cm} C{1.1cm} | C{1.1cm} C{1.1cm} C{1.1cm} C{1.1cm} C{1.1cm} }
    \ChangeRT{1pt} &
    \multicolumn{5}{c}{\textbf{Web Search (P)}} & \multicolumn{5}{|c}{\textbf{Web Search (D)}}  \\
    \cmidrule(lr){1-1}
    \cmidrule(lr){2-6}
    \cmidrule(lr){7-11}
    \textbf{Methods} & 
    \textbf{R@10} & \textbf{R@20} & \textbf{R@30} & \textbf{R@50} & \textbf{R@100} & \textbf{R@10} & \textbf{R@20} & \textbf{R@30} & \textbf{R@50} & \textbf{R@100}  \\
    \hline
    DPR + HNSW & 0.5328 & 0.6413 & 0.6984 & 0.7653 & 0.8405 & 0.5983 & 0.7059 & 0.7562 & 0.8110 & 0.8688  \\
    STAR + HNSW & \underline{0.5919} & \underline{0.6917} & \underline{0.7468} & \underline{0.8038} & \underline{0.8611} & \underline{0.6816} & \underline{0.7775} & \underline{0.8236} & \underline{0.8648} & \underline{0.9044}  \\
    ANCE + HNSW & 0.5842 & 0.6858 & 0.7427 & 0.7991 & 0.8596 & 0.6728 & 0.7633 & 0.8060 & 0.8476 & 0.8840  \\
    DE-BERT + HNSW & 0.5540 & 0.6580 & 0.7121 & 0.7709 & 0.8350 & 0.6112 & 0.7142 & 0.7660 & 0.8195 & 0.8707  \\
    \hline
    \textbf{BiDR (RD)} & 0.6048 & 0.7098 & 0.7604 & 0.8156 & 0.8753 & \textbf{0.6988} & \textbf{0.7964} & \textbf{0.8399} & \textbf{0.8821} & 0.9176 \\
    \textbf{BiDR (SN)} & \textbf{0.6087} & \textbf{0.7116} & \textbf{0.7604} & \textbf{0.8169} & \textbf{0.8786} & 0.6959 & 0.7924 & 0.8357 & 0.8808 & \textbf{0.9225} \\
    \ChangeRT{1pt}
    \end{tabular}
    \caption{\small Evaluation of generic EBR. BiDR (RD) and (SN) maintain their leading performances after baselines' switching to HNSW.}
    \vspace{-23pt}
    \label{tab:2}
\end{table*}


\vspace{-10pt}
\subsubsection{Evaluation of generic EBR}\label{sec:exp-2}
In this part, the baseline ANN index is switched to HNSW, which is more accurate than DiskANN but no longer memory efficient.
Given that the Sponsored Search dataset has a billion-scale corpus, which is too large for HNSW to fit into memory, we leverage Web Search (P) and (D) for the evaluation.
The experiment results are shown as Table \ref{tab:2}: with the utilization of HNSW, the baselines' performances are improved compared with their previous results in Table \ref{tab:1}. At the same time, BiDR (RD) and (SN) still maintain their advantages over the strongest baselines; e.g., \textbf{Recall@10} can be relatively improved by \textbf{2.66\%} and \textbf{2.52\%} on different datasets. {Such a finding indicates our bi-granular embeddings' competitiveness against the conventional EBR methods, which rely on a unified set of dense embeddings.} The explanation about our advantage is made as follows. \textit{The global discrimination and local discrimination are two different objectives, whose effects can be contradicted}: a too strong local discrimination is unfavorable to the global discrimination, and vice versa. Therefore, instead of enforcing a unified set of embeddings, it's better to leverage two sets of embeddings which work collaboratively to optimize each of the objectives. This point will be further clarified by {our experiments on sampling strategies in \ref{sec:exp-4}}. Besides, it should also be noted that the maintenance of bi-granular embeddings is not expensive at all: for one thing, the bi-granular embeddings are inferred offline; for the other thing, the sparse embeddings are highly memory and search efficient, where little extra cost is introduced compared with the method which merely uses the dense embeddings.

\subsubsection{{Efficiency and Scalability}}\label{sec:exp-3}
\begin{figure}[t]
\centering
\includegraphics[width=0.95\linewidth]{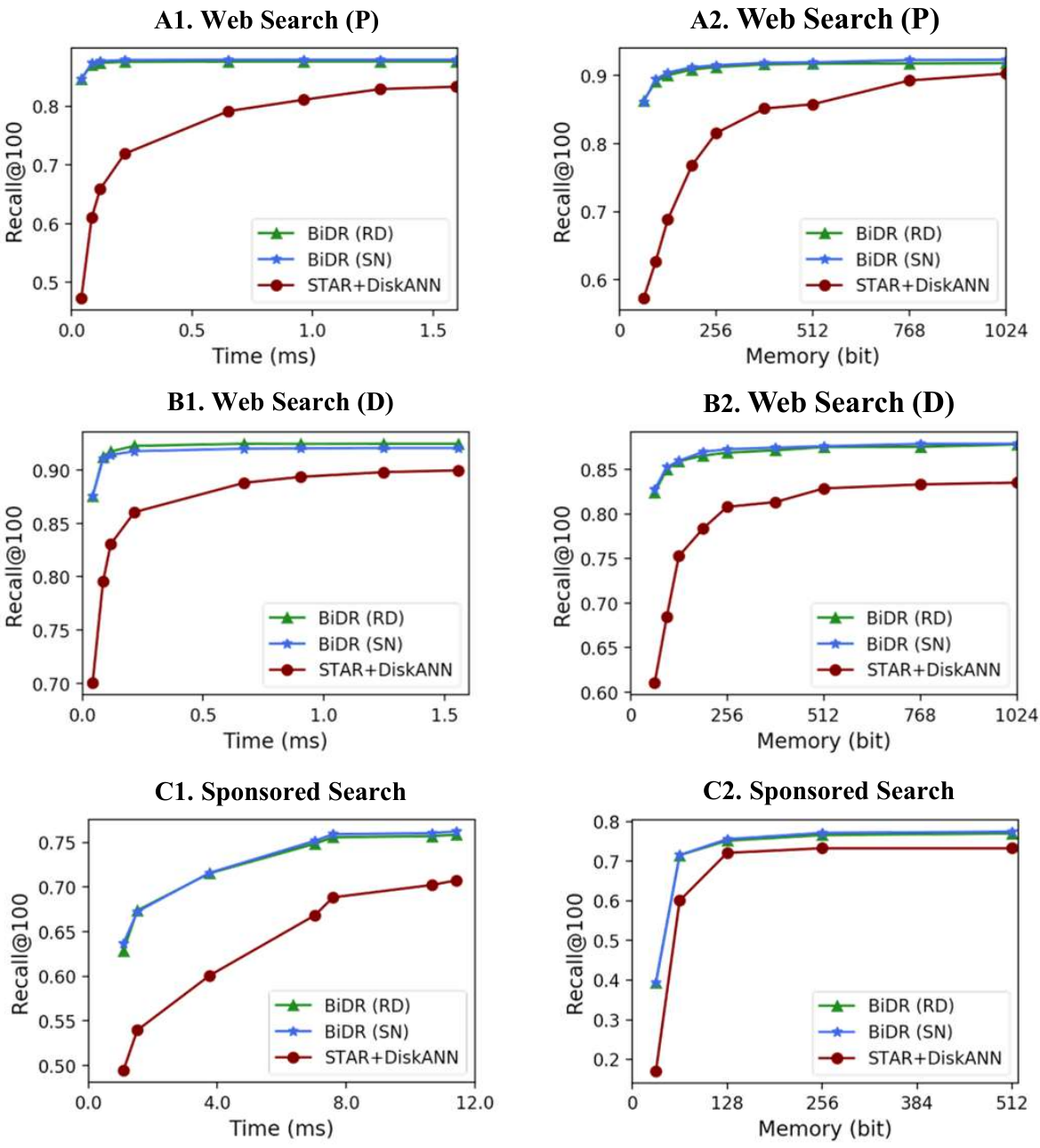}
\vspace{-12pt}
\caption{\small {Evaluation of efficiency ($\#1$) and scalability ($\#2$). BiDR reaches higher recall rates with less latency and memory cost.}}
\vspace{-3pt}
\label{fig:4}
\end{figure}

The \textbf{evaluation of efficiency} is shown as Figure \ref{fig:4}: \textbf{A1}/\textbf{B1} for Web Search (P)/(D), \textbf{C1} for Sponsored Search. In our experiment, the time latency and recall rate are measured for different scales of candidates. On one hand, the expanded candidate size will contribute to a higher recall rate, as the ground-truth answers are more likely to be included (the ANN search will be reduced to linear scan when the entire corpus is included). On the other hand, the expanded candidates will increase the cost of search and post verification, which incurs a higher time latency. In our experiment, the candidate size is increased from 100 to 10000, which leads to the growth of recall rate as well as time latency. For Web Search (P) and (D): we observe that the \textbf{Recall@100} about BiDR (RD) and (SN) soon reach a highly competitive level (less than \textbf{0.1 ms}, with only around \textbf{300} candidates included), which indicates the time efficiency of our method. Compared with A1 and B1, it takes more latency for BiDR to reach a stable competitive Recall@100 on Sponsored Search, as shown in C1 (about \textbf{8ms}, with \textbf{5000} candidates retrieved and re-ranked). Considering that Sponsored Search has a billion-scale corpus (over $\times$100 larger than the Web Search datasets), the increase of search cost can be expected. Besides, the observed latency is indeed competitive in reality, as it is measured on the basis of one physical machine (detailed specification is made in Appendix). By comparison, ``STAR + DiskANN'' (the strongest baseline) is limited when the candidate size is small; this is mainly caused by the lossy unsupervised quantization of DiskANN. Besides, the performance gap between BiDR and ``STAR + DiskANN'' still remains when huge amounts of candidates are included (\textbf{10000}); this is resulted from the superiority of our progressively optimized bi-granular embeddings, as analyzed in \ref{sec:exp-2}.

The scalability is evaluated as Figure \ref{fig:4} $\{A2,B2,C2\}$. The scale of sparse embeddings is gradually increased by expanding the codebooks, whose memory cost grows from 64 to 1024 bit. The data quantization will be less lossy when more bits are allocated to the sparse embeddings; thus, the recall rate can be improved. For all datasets: BiDR soon grows to high recall rates with less than \textbf{256 bit} (per answer) memory cost, which means an one-billion corpus can be hosted for high-quality ANN search with merely \textbf{32 GB} RAM cost. 
Such a finding reflects our capability of handling massive-scale EBR scenarios. Similar to the observations made in efficiency analysis: 1) for Sponsored Search, all methods suffer from low recall accuracy at small bit rates; and 2) DiskANN+STAR's recall rates are limited at small bit rates, and the performance gap remains at high bit rates. We attribute these observations to the same reasons as explained in our efficiency analysis.

\begin{table*}[t]
    \centering
    \footnotesize
    \begin{tabular}{p{2.0cm} | C{1.2cm} C{1.2cm} C{1.2cm} | C{1.2cm} C{1.2cm} C{1.2cm} | C{1.2cm} C{1.2cm} C{1.2cm} }
    \ChangeRT{1pt} &
    \multicolumn{3}{c}{\textbf{Web Search (P)}} & \multicolumn{3}{|c}{\textbf{Web Search (D)}} & 
    \multicolumn{3}{|c}{\textbf{Sponsored Search}} \\
    \cmidrule(lr){1-1}
    \cmidrule(lr){2-4}
    \cmidrule(lr){5-7}
    \cmidrule(lr){8-10}
    \textbf{Quantization} & 
    \textbf{R@50} & \textbf{R@100} & \textbf{R@1000} & \textbf{R@50} & \textbf{R@100} & \textbf{R@1000} & \textbf{R@50} & \textbf{R@100} & \textbf{R@1000} \\
    \hline
    PQ  & 0.7127 & 0.7867 & 0.9349 & 0.4032 & 0.4900 & 0.7488 & 0.2413 & 0.3167 & 0.6000 \\
    OPQ  & 0.7598 & 0.8270 & 0.9519 & 0.7864 & 0.8490 & 0.9483 & 0.2629 & 0.3320 & 0.6140 \\
    DPQ  & 0.7696 & 0.8399 & 0.9558 & 0.7993 & 0.8571 & 0.9543 & 0.3359 & 0.4288 & 0.7218 \\
    Contrastive & \textbf{0.7801} & \textbf{0.8460} & \textbf{0.9605} & \textbf{0.8182} & \textbf{0.8702} & \textbf{0.9612} & \textbf{0.4067} & \textbf{0.5004} & \textbf{0.7659} \\
    \hline \hline
    \textbf{Sampling} & 
    \textbf{R@10} & \textbf{R@50} & \textbf{R@100} & \textbf{R@10} & \textbf{R@50} & \textbf{R@100} & \textbf{R@10} & \textbf{R@50} & \textbf{R@100}  \\
    \hline
    ANN & 0.6021 & 0.8101 & 0.8728 & 0.6921 & 0.8709 & 0.9109 & 0.3835 & 0.6302 & 0.7041  \\
    ANN + In-batch & 0.6014 & 0.8092 & 0.8727 & 0.6861 & 0.8698 & 0.9107 & 0.3755 & 0.6254 & 0.6999 \\
    Random Walk & 0.6048 & 0.8156 & 0.8753 & \textbf{0.6988} & \textbf{0.8821} & \textbf{0.9176} & 0.3985 & 0.6467 & 0.7152 \\
    Snowball & \textbf{0.6087} & \textbf{0.8169} & \textbf{0.8786} & 0.6959 & 0.8808 & 0.9225  & \textbf{0.4038} & \textbf{0.6489} & \textbf{0.7158} \\
    \hline \hline
    \textbf{Unified Query} & 
    \textbf{R@10} & \textbf{R@50} & \textbf{R@100} & \textbf{R@10} & \textbf{R@50} & \textbf{R@100} & \textbf{R@10} & \textbf{R@50} & \textbf{R@100}  \\
    \hline
    Unified (RD) & 0.6027 & 0.8121 & 0.8663 & \textbf{0.7003} & 0.8785 & 0.9147  & 0.3962 & 0.6346 & 0.7046 \\
    w.o. Unified (RD) & 0.6048 & 0.8156 & 0.8753 & 0.6988 & \textbf{0.8821} & 0.9176 & 0.3985 & 0.6467 & 0.7152 \\
    Unified (SN) & 0.6003 & 0.8098 & 0.8704 & 0.6924 & 0.8775 & 0.9143 & 0.4002 & 0.6399 & 0.7064 \\
    w.o. Unified (SN) & \textbf{0.6087} & \textbf{0.8169} & \textbf{0.8786} & 0.6959 & 0.8808 & \textbf{0.9225}  & \textbf{0.4038} & \textbf{0.6489} & \textbf{0.7158} \\
    \ChangeRT{1pt}
    \end{tabular}
    \caption{\small Ablation studies: the impact of (1) contrastive quantization, (2) locality-centric sampling (LCS) and (3) query unification. {(Recall@$\mathbf{1000}$ is reported for the quantization methods, considering that the Top-1000 answers are selected in candidate search.)}}
    \vspace{-22pt}
    \label{tab:3}
\end{table*}

\subsubsection{Ablation Studies}\label{sec:exp-4}
The detailed impact is investigated for contrastive quantization, locality-centric sampling, and unified query.

\textbf{Quantization}. The evaluation of our \textit{contrastive} quantization is shown in the upper part of Table \ref{tab:3}, with three representative baselines compared: 1) the typical unsupervised method \textit{PQ} \cite{jegou2010product}; 2) \textit{OPQ} \cite{ge2013optimized}, which iteratively learns the embedding's projection and codebooks to minimize the reconstruction loss; 3) \textit{DPQ} \cite{chen2020differentiable}, where the reconstruction loss and matching loss are jointly minimized within one network. {The recall rates are reported for the candidate search result, whose size is much larger than the final retrieval result.} We may find that the direct use of PQ is lossy, whose accuracy is the lowest of all. Other baselines are improved thanks to the effective reduction of reconstruction loss; however, they are still inferior to our contrastive quantization, where the InfoNCE loss (the objective for retrieval accuracy) is minimized for the sparse embeddings.
{In fact, we find that the gap in recall rate between our contrastive quantization and the continuous upper bound (where dense embeddings are used without quantization) can be even smaller than $1e$-3 for the Top-1000 candidates (the default size of candidates). Such a tiny loss is almost ignorable in practice.}


\textbf{Sampling}. The impacts of locality-centric sampling methods (LCS): \textit{Random Walk} and \textit{Snowball} Sampling, are evaluated against two baseline strategies: 1) \textit{ANN} negative sampling as used by ANCE \cite{xiong2021approximate}, where each query is assigned with a fixed amount of negatives sampled by ANN; 2) \textit{ANN + In-batch} as used by STAR \cite{zhan2021optimizing}, where the ANN negatives are combined with in-batch negatives. The evaluation results are shown as the middle part of Table \ref{tab:3}, in which both Random Walk and Snowball outperform the baseline sampling strategies. Such an observation verifies that LCS is beneficial to the learning of dense embeddings, where the highly concentrated in-batch negatives may provide strong supervision signals for the local discrimination. Besides, as discussed in \ref{sec:exp-1}, Random Walk and Snowball take the lead under different conditions. For Web Search (P) and Sponsored Search where a great deal of consecutive LCS samplings are needed, Snowball achieves higher accuracy because of the better preservation of locality. For Web Search (D) where LCS sampling steps are limited to be small, Random Walk performs better as more diversified samples can be introduced.


\textbf{One interesting observation to emphasize}: although STAR is generally better than ANCE (Table \ref{tab:1} and \ref{tab:2}), {its sampling strategy (ANN + In-batch) is worse than the one used by ANCE (ANN) when applied to our dense embeddings' learning.}
A fundamental difference between the two strategies is that the ANN negatives focus on the local discrimination, while the ANN + In-batch negatives aim to reach a trade-off between the local and global discrimination. Knowing the dense embeddings are learned for post verification, where the local discrimination matters the most, such a trade-off is undesirable and leads to a negative effect on retrieval accuracy. This observation echoes our analysis in \ref{sec:exp-2}, which indicates that the effects of local and global discrimination are not always consistent. On top of our progressively optimized bi-granular document representation, \textit{i.e.}, with the sparse embeddings optimized for the global discrimination and the dense embeddings optimized for local discrimination, the retrieval accuracy can be effectively enhanced. 

\textbf{Query Unification}. We analyze the impact of unifying the query encoder (Table \ref{tab:3}). It is observed that the loss from query unification is small; e.g., the gaps between the non-unified results are merely \textbf{-0.0015} $\sim$ \textbf{0.0084} (\textbf{0.0031} on average) in terms of \textbf{Recall@10}. {Such a performance loss is far smaller than BiDR's advantages over the baselines (shown in Table \ref{tab:1} and \ref{tab:2}), which means the query unification may well preserve the retrieval accuracy while cutting down BiDR's online serving cost.} Meanwhile, given that the performance of ``w.o. Unified'' is still slightly better in general, people may keep the non-unified query encoders running in parallel when the system's capacity is abundant.

\subsubsection{Online Performance}
BiDR has been successfully applied to a major sponsored search platform in the world, which retrieves ads in response to user's search queries. The previous system already adopted an ensemble of competitive scalable EBR solutions, such as BERT based sentence embeddings plus DiskANN, whose performances are on par with the strongest baselines in our offline studies. Even with such a strong comparison group, BiDR achieves notable improvements for the key performance indicators: there have been \textbf{+1.95\% RPM gain} (revenue per mille, i.e., per thousand impressions), \textbf{+1.01\% Recall gain}, and \textbf{+0.49\% CTR gain},  which means substantial improvements on profit and user experience. 

\section{Conclusion}
{In this work, we proposed the progressively optimized bi-granular document representation for scalable embedding based retrieval. 
The contrastive quantization was used to generate the sparse embeddings, which led to precise and memory-efficient candidate search. Conditioned on the candidate distribution, the dense embeddings were continuously learned with locality-centric sampling, which achieved superior post verification accuracy. Comprehensive offline and online evaluations were performed, whose results verified the effectiveness, efficiency and scalability of our method.}

\bibliographystyle{ACM-Reference-Format}
\bibliography{ref}


\begin{thebibliography}{40}


\ifx \showCODEN    \undefined \def \showCODEN     #1{\unskip}     \fi
\ifx \showDOI      \undefined \def \showDOI       #1{#1}\fi
\ifx \showISBNx    \undefined \def \showISBNx     #1{\unskip}     \fi
\ifx \showISBNxiii \undefined \def \showISBNxiii  #1{\unskip}     \fi
\ifx \showISSN     \undefined \def \showISSN      #1{\unskip}     \fi
\ifx \showLCCN     \undefined \def \showLCCN      #1{\unskip}     \fi
\ifx \shownote     \undefined \def \shownote      #1{#1}          \fi
\ifx \showarticletitle \undefined \def \showarticletitle #1{#1}   \fi
\ifx \showURL      \undefined \def \showURL       {\relax}        \fi
\providecommand\bibfield[2]{#2}
\providecommand\bibinfo[2]{#2}
\providecommand\natexlab[1]{#1}
\providecommand\showeprint[2][]{arXiv:#2}

\bibitem[\protect\citeauthoryear{Baranchuk, Babenko, and Malkov}{Baranchuk
  et~al\mbox{.}}{2018}]%
        {baranchuk2018revisiting}
\bibfield{author}{\bibinfo{person}{Dmitry Baranchuk}, \bibinfo{person}{Artem
  Babenko}, {and} \bibinfo{person}{Yury Malkov}.}
  \bibinfo{year}{2018}\natexlab{}.
\newblock \showarticletitle{Revisiting the inverted indices for billion-scale
  approximate nearest neighbors}. In \bibinfo{booktitle}{\emph{Proceedings of
  the European Conference on Computer Vision (ECCV)}}.
  \bibinfo{pages}{202--216}.
\newblock


\bibitem[\protect\citeauthoryear{Beis and Lowe}{Beis and Lowe}{1997}]%
        {beis1997shape}
\bibfield{author}{\bibinfo{person}{Jeffrey~S Beis} {and}
  \bibinfo{person}{David~G Lowe}.} \bibinfo{year}{1997}\natexlab{}.
\newblock \showarticletitle{Shape indexing using approximate nearest-neighbour
  search in high-dimensional spaces}. In \bibinfo{booktitle}{\emph{Proceedings
  of IEEE computer society conference on computer vision and pattern
  recognition}}. IEEE, \bibinfo{pages}{1000--1006}.
\newblock


\bibitem[\protect\citeauthoryear{Bengio, L{\'e}onard, and Courville}{Bengio
  et~al\mbox{.}}{2013}]%
        {bengio2013estimating}
\bibfield{author}{\bibinfo{person}{Yoshua Bengio}, \bibinfo{person}{Nicholas
  L{\'e}onard}, {and} \bibinfo{person}{Aaron Courville}.}
  \bibinfo{year}{2013}\natexlab{}.
\newblock \showarticletitle{Estimating or propagating gradients through
  stochastic neurons for conditional computation}.
\newblock \bibinfo{journal}{\emph{arXiv preprint arXiv:1308.3432}}
  (\bibinfo{year}{2013}).
\newblock


\bibitem[\protect\citeauthoryear{Chang, Yu, Chang, Yang, and Kumar}{Chang
  et~al\mbox{.}}{2020}]%
        {chang2020pre}
\bibfield{author}{\bibinfo{person}{Wei-Cheng Chang}, \bibinfo{person}{Felix~X
  Yu}, \bibinfo{person}{Yin-Wen Chang}, \bibinfo{person}{Yiming Yang}, {and}
  \bibinfo{person}{Sanjiv Kumar}.} \bibinfo{year}{2020}\natexlab{}.
\newblock \showarticletitle{Pre-training tasks for embedding-based large-scale
  retrieval}.
\newblock \bibinfo{journal}{\emph{arXiv preprint arXiv:2002.03932}}
  (\bibinfo{year}{2020}).
\newblock


\bibitem[\protect\citeauthoryear{Chen, Kornblith, Norouzi, and Hinton}{Chen
  et~al\mbox{.}}{2020a}]%
        {chen2020simple}
\bibfield{author}{\bibinfo{person}{Ting Chen}, \bibinfo{person}{Simon
  Kornblith}, \bibinfo{person}{Mohammad Norouzi}, {and}
  \bibinfo{person}{Geoffrey Hinton}.} \bibinfo{year}{2020}\natexlab{a}.
\newblock \showarticletitle{A simple framework for contrastive learning of
  visual representations}. In \bibinfo{booktitle}{\emph{International
  conference on machine learning}}. PMLR, \bibinfo{pages}{1597--1607}.
\newblock


\bibitem[\protect\citeauthoryear{Chen, Li, and Sun}{Chen
  et~al\mbox{.}}{2020b}]%
        {chen2020differentiable}
\bibfield{author}{\bibinfo{person}{Ting Chen}, \bibinfo{person}{Lala Li}, {and}
  \bibinfo{person}{Yizhou Sun}.} \bibinfo{year}{2020}\natexlab{b}.
\newblock \showarticletitle{Differentiable product quantization for end-to-end
  embedding compression}. In \bibinfo{booktitle}{\emph{International Conference
  on Machine Learning}}. PMLR, \bibinfo{pages}{1617--1626}.
\newblock


\bibitem[\protect\citeauthoryear{Craswell, Mitra, Yilmaz, Campos, and
  Voorhees}{Craswell et~al\mbox{.}}{2020}]%
        {craswell2020overview}
\bibfield{author}{\bibinfo{person}{Nick Craswell}, \bibinfo{person}{Bhaskar
  Mitra}, \bibinfo{person}{Emine Yilmaz}, \bibinfo{person}{Daniel Campos},
  {and} \bibinfo{person}{Ellen~M Voorhees}.} \bibinfo{year}{2020}\natexlab{}.
\newblock \showarticletitle{Overview of the trec 2019 deep learning track}.
\newblock \bibinfo{journal}{\emph{arXiv preprint arXiv:2003.07820}}
  (\bibinfo{year}{2020}).
\newblock


\bibitem[\protect\citeauthoryear{Datar, Immorlica, Indyk, and Mirrokni}{Datar
  et~al\mbox{.}}{2004}]%
        {datar2004locality}
\bibfield{author}{\bibinfo{person}{Mayur Datar}, \bibinfo{person}{Nicole
  Immorlica}, \bibinfo{person}{Piotr Indyk}, {and} \bibinfo{person}{Vahab~S
  Mirrokni}.} \bibinfo{year}{2004}\natexlab{}.
\newblock \showarticletitle{Locality-sensitive hashing scheme based on p-stable
  distributions}. In \bibinfo{booktitle}{\emph{Proceedings of the twentieth
  annual symposium on Computational geometry}}. \bibinfo{pages}{253--262}.
\newblock


\bibitem[\protect\citeauthoryear{Devlin, Chang, Lee, and Toutanova}{Devlin
  et~al\mbox{.}}{2018}]%
        {devlin2018bert}
\bibfield{author}{\bibinfo{person}{Jacob Devlin}, \bibinfo{person}{Ming-Wei
  Chang}, \bibinfo{person}{Kenton Lee}, {and} \bibinfo{person}{Kristina
  Toutanova}.} \bibinfo{year}{2018}\natexlab{}.
\newblock \showarticletitle{Bert: Pre-training of deep bidirectional
  transformers for language understanding}.
\newblock \bibinfo{journal}{\emph{arXiv preprint arXiv:1810.04805}}
  (\bibinfo{year}{2018}).
\newblock


\bibitem[\protect\citeauthoryear{Fan, Guo, Zhu, Miao, Sun, and Li}{Fan
  et~al\mbox{.}}{2019}]%
        {fan2019mobius}
\bibfield{author}{\bibinfo{person}{Miao Fan}, \bibinfo{person}{Jiacheng Guo},
  \bibinfo{person}{Shuai Zhu}, \bibinfo{person}{Shuo Miao},
  \bibinfo{person}{Mingming Sun}, {and} \bibinfo{person}{Ping Li}.}
  \bibinfo{year}{2019}\natexlab{}.
\newblock \showarticletitle{MOBIUS: towards the next generation of query-ad
  matching in baidu's sponsored search}. In
  \bibinfo{booktitle}{\emph{Proceedings of the 25th ACM SIGKDD International
  Conference on Knowledge Discovery \& Data Mining}}.
  \bibinfo{pages}{2509--2517}.
\newblock


\bibitem[\protect\citeauthoryear{Fu, Xiang, Wang, and Cai}{Fu
  et~al\mbox{.}}{2017}]%
        {fu2017fast}
\bibfield{author}{\bibinfo{person}{Cong Fu}, \bibinfo{person}{Chao Xiang},
  \bibinfo{person}{Changxu Wang}, {and} \bibinfo{person}{Deng Cai}.}
  \bibinfo{year}{2017}\natexlab{}.
\newblock \showarticletitle{Fast approximate nearest neighbor search with the
  navigating spreading-out graph}.
\newblock \bibinfo{journal}{\emph{arXiv preprint arXiv:1707.00143}}
  (\bibinfo{year}{2017}).
\newblock


\bibitem[\protect\citeauthoryear{Gao, Yao, and Chen}{Gao et~al\mbox{.}}{2021}]%
        {gao2021simcse}
\bibfield{author}{\bibinfo{person}{Tianyu Gao}, \bibinfo{person}{Xingcheng
  Yao}, {and} \bibinfo{person}{Danqi Chen}.} \bibinfo{year}{2021}\natexlab{}.
\newblock \showarticletitle{SimCSE: Simple Contrastive Learning of Sentence
  Embeddings}.
\newblock \bibinfo{journal}{\emph{arXiv preprint arXiv:2104.08821}}
  (\bibinfo{year}{2021}).
\newblock


\bibitem[\protect\citeauthoryear{Ge, He, Ke, and Sun}{Ge et~al\mbox{.}}{2013}]%
        {ge2013optimized}
\bibfield{author}{\bibinfo{person}{Tiezheng Ge}, \bibinfo{person}{Kaiming He},
  \bibinfo{person}{Qifa Ke}, {and} \bibinfo{person}{Jian Sun}.}
  \bibinfo{year}{2013}\natexlab{}.
\newblock \showarticletitle{Optimized product quantization}.
\newblock \bibinfo{journal}{\emph{IEEE transactions on pattern analysis and
  machine intelligence}} \bibinfo{volume}{36}, \bibinfo{number}{4}
  (\bibinfo{year}{2013}), \bibinfo{pages}{744--755}.
\newblock


\bibitem[\protect\citeauthoryear{Guu, Lee, Tung, Pasupat, and Chang}{Guu
  et~al\mbox{.}}{2020}]%
        {guu2020realm}
\bibfield{author}{\bibinfo{person}{Kelvin Guu}, \bibinfo{person}{Kenton Lee},
  \bibinfo{person}{Zora Tung}, \bibinfo{person}{Panupong Pasupat}, {and}
  \bibinfo{person}{Ming-Wei Chang}.} \bibinfo{year}{2020}\natexlab{}.
\newblock \showarticletitle{Realm: Retrieval-augmented language model
  pre-training}.
\newblock \bibinfo{journal}{\emph{arXiv preprint arXiv:2002.08909}}
  (\bibinfo{year}{2020}).
\newblock


\bibitem[\protect\citeauthoryear{He, Fan, Wu, Xie, and Girshick}{He
  et~al\mbox{.}}{2020}]%
        {he2020momentum}
\bibfield{author}{\bibinfo{person}{Kaiming He}, \bibinfo{person}{Haoqi Fan},
  \bibinfo{person}{Yuxin Wu}, \bibinfo{person}{Saining Xie}, {and}
  \bibinfo{person}{Ross Girshick}.} \bibinfo{year}{2020}\natexlab{}.
\newblock \showarticletitle{Momentum contrast for unsupervised visual
  representation learning}. In \bibinfo{booktitle}{\emph{Proceedings of the
  IEEE/CVF Conference on Computer Vision and Pattern Recognition}}.
  \bibinfo{pages}{9729--9738}.
\newblock


\bibitem[\protect\citeauthoryear{Huang, Sharma, Sun, Xia, Zhang, Pronin,
  Padmanabhan, Ottaviano, and Yang}{Huang et~al\mbox{.}}{2020}]%
        {huang2020embedding}
\bibfield{author}{\bibinfo{person}{Jui-Ting Huang}, \bibinfo{person}{Ashish
  Sharma}, \bibinfo{person}{Shuying Sun}, \bibinfo{person}{Li Xia},
  \bibinfo{person}{David Zhang}, \bibinfo{person}{Philip Pronin},
  \bibinfo{person}{Janani Padmanabhan}, \bibinfo{person}{Giuseppe Ottaviano},
  {and} \bibinfo{person}{Linjun Yang}.} \bibinfo{year}{2020}\natexlab{}.
\newblock \showarticletitle{Embedding-based retrieval in facebook search}. In
  \bibinfo{booktitle}{\emph{Proceedings of the 26th ACM SIGKDD International
  Conference on Knowledge Discovery \& Data Mining}}.
  \bibinfo{pages}{2553--2561}.
\newblock


\bibitem[\protect\citeauthoryear{Jegou, Douze, and Schmid}{Jegou
  et~al\mbox{.}}{2010}]%
        {jegou2010product}
\bibfield{author}{\bibinfo{person}{Herve Jegou}, \bibinfo{person}{Matthijs
  Douze}, {and} \bibinfo{person}{Cordelia Schmid}.}
  \bibinfo{year}{2010}\natexlab{}.
\newblock \showarticletitle{Product quantization for nearest neighbor search}.
\newblock \bibinfo{journal}{\emph{IEEE transactions on pattern analysis and
  machine intelligence}} \bibinfo{volume}{33}, \bibinfo{number}{1}
  (\bibinfo{year}{2010}), \bibinfo{pages}{117--128}.
\newblock


\bibitem[\protect\citeauthoryear{J{\'e}gou, Tavenard, Douze, and
  Amsaleg}{J{\'e}gou et~al\mbox{.}}{2011}]%
        {jegou2011searching}
\bibfield{author}{\bibinfo{person}{Herv{\'e} J{\'e}gou},
  \bibinfo{person}{Romain Tavenard}, \bibinfo{person}{Matthijs Douze}, {and}
  \bibinfo{person}{Laurent Amsaleg}.} \bibinfo{year}{2011}\natexlab{}.
\newblock \showarticletitle{Searching in one billion vectors: re-rank with
  source coding}. In \bibinfo{booktitle}{\emph{2011 IEEE International
  Conference on Acoustics, Speech and Signal Processing (ICASSP)}}. IEEE,
  \bibinfo{pages}{861--864}.
\newblock


\bibitem[\protect\citeauthoryear{Joshi, Chen, Liu, Weld, Zettlemoyer, and
  Levy}{Joshi et~al\mbox{.}}{2020}]%
        {joshi2020spanbert}
\bibfield{author}{\bibinfo{person}{Mandar Joshi}, \bibinfo{person}{Danqi Chen},
  \bibinfo{person}{Yinhan Liu}, \bibinfo{person}{Daniel~S Weld},
  \bibinfo{person}{Luke Zettlemoyer}, {and} \bibinfo{person}{Omer Levy}.}
  \bibinfo{year}{2020}\natexlab{}.
\newblock \showarticletitle{Spanbert: Improving pre-training by representing
  and predicting spans}.
\newblock \bibinfo{journal}{\emph{Transactions of the Association for
  Computational Linguistics}}  \bibinfo{volume}{8} (\bibinfo{year}{2020}),
  \bibinfo{pages}{64--77}.
\newblock


\bibitem[\protect\citeauthoryear{Karpukhin, Oguz, Min, Lewis, Wu, Edunov, Chen,
  and Yih}{Karpukhin et~al\mbox{.}}{2020}]%
        {karpukhin2020dense}
\bibfield{author}{\bibinfo{person}{Vladimir Karpukhin}, \bibinfo{person}{Barlas
  Oguz}, \bibinfo{person}{Sewon Min}, \bibinfo{person}{Patrick Lewis},
  \bibinfo{person}{Ledell Wu}, \bibinfo{person}{Sergey Edunov},
  \bibinfo{person}{Danqi Chen}, {and} \bibinfo{person}{Wen-tau Yih}.}
  \bibinfo{year}{2020}\natexlab{}.
\newblock \showarticletitle{Dense Passage Retrieval for Open-Domain Question
  Answering}. In \bibinfo{booktitle}{\emph{Proceedings of the 2020 Conference
  on Empirical Methods in Natural Language Processing (EMNLP)}}.
  \bibinfo{pages}{6769--6781}.
\newblock


\bibitem[\protect\citeauthoryear{Le and Mikolov}{Le and Mikolov}{2014}]%
        {le2014distributed}
\bibfield{author}{\bibinfo{person}{Quoc Le} {and} \bibinfo{person}{Tomas
  Mikolov}.} \bibinfo{year}{2014}\natexlab{}.
\newblock \showarticletitle{Distributed representations of sentences and
  documents}. In \bibinfo{booktitle}{\emph{International conference on machine
  learning}}. PMLR, \bibinfo{pages}{1188--1196}.
\newblock


\bibitem[\protect\citeauthoryear{Lee, Chang, and Toutanova}{Lee
  et~al\mbox{.}}{2019}]%
        {lee2019latent}
\bibfield{author}{\bibinfo{person}{Kenton Lee}, \bibinfo{person}{Ming-Wei
  Chang}, {and} \bibinfo{person}{Kristina Toutanova}.}
  \bibinfo{year}{2019}\natexlab{}.
\newblock \showarticletitle{Latent retrieval for weakly supervised open domain
  question answering}.
\newblock \bibinfo{journal}{\emph{arXiv preprint arXiv:1906.00300}}
  (\bibinfo{year}{2019}).
\newblock


\bibitem[\protect\citeauthoryear{Lin, Yang, and Lin}{Lin et~al\mbox{.}}{2020}]%
        {lin2020distilling}
\bibfield{author}{\bibinfo{person}{Sheng-Chieh Lin},
  \bibinfo{person}{Jheng-Hong Yang}, {and} \bibinfo{person}{Jimmy Lin}.}
  \bibinfo{year}{2020}\natexlab{}.
\newblock \showarticletitle{Distilling dense representations for ranking using
  tightly-coupled teachers}.
\newblock \bibinfo{journal}{\emph{arXiv preprint arXiv:2010.11386}}
  (\bibinfo{year}{2020}).
\newblock


\bibitem[\protect\citeauthoryear{Liu, Lu, Cheng, Shi, Wang, Cheng, and Yin}{Liu
  et~al\mbox{.}}{2021}]%
        {liu2021pre}
\bibfield{author}{\bibinfo{person}{Yiding Liu}, \bibinfo{person}{Weixue Lu},
  \bibinfo{person}{Suqi Cheng}, \bibinfo{person}{Daiting Shi},
  \bibinfo{person}{Shuaiqiang Wang}, \bibinfo{person}{Zhicong Cheng}, {and}
  \bibinfo{person}{Dawei Yin}.} \bibinfo{year}{2021}\natexlab{}.
\newblock \showarticletitle{Pre-trained Language Model for Web-scale Retrieval
  in Baidu Search}. In \bibinfo{booktitle}{\emph{Proceedings of the 27th ACM
  SIGKDD Conference on Knowledge Discovery and Data Mining}}.
  \bibinfo{pages}{3365--3375}.
\newblock


\bibitem[\protect\citeauthoryear{Liu, Ott, Goyal, Du, Joshi, Chen, Levy, Lewis,
  Zettlemoyer, and Stoyanov}{Liu et~al\mbox{.}}{2019}]%
        {liu2019roberta}
\bibfield{author}{\bibinfo{person}{Yinhan Liu}, \bibinfo{person}{Myle Ott},
  \bibinfo{person}{Naman Goyal}, \bibinfo{person}{Jingfei Du},
  \bibinfo{person}{Mandar Joshi}, \bibinfo{person}{Danqi Chen},
  \bibinfo{person}{Omer Levy}, \bibinfo{person}{Mike Lewis},
  \bibinfo{person}{Luke Zettlemoyer}, {and} \bibinfo{person}{Veselin
  Stoyanov}.} \bibinfo{year}{2019}\natexlab{}.
\newblock \showarticletitle{Roberta: A robustly optimized bert pretraining
  approach}.
\newblock \bibinfo{journal}{\emph{arXiv preprint arXiv:1907.11692}}
  (\bibinfo{year}{2019}).
\newblock


\bibitem[\protect\citeauthoryear{Lu, Jiao, and Zhang}{Lu et~al\mbox{.}}{2020}]%
        {lu2020twinbert}
\bibfield{author}{\bibinfo{person}{Wenhao Lu}, \bibinfo{person}{Jian Jiao},
  {and} \bibinfo{person}{Ruofei Zhang}.} \bibinfo{year}{2020}\natexlab{}.
\newblock \showarticletitle{Twinbert: Distilling knowledge to twin-structured
  compressed BERT models for large-scale retrieval}. In
  \bibinfo{booktitle}{\emph{Proceedings of the 29th ACM International
  Conference on Information \& Knowledge Management}}.
  \bibinfo{pages}{2645--2652}.
\newblock


\bibitem[\protect\citeauthoryear{Luan, Eisenstein, Toutanova, and Collins}{Luan
  et~al\mbox{.}}{2021}]%
        {luan2021sparse}
\bibfield{author}{\bibinfo{person}{Yi Luan}, \bibinfo{person}{Jacob
  Eisenstein}, \bibinfo{person}{Kristina Toutanova}, {and}
  \bibinfo{person}{Michael Collins}.} \bibinfo{year}{2021}\natexlab{}.
\newblock \showarticletitle{Sparse, Dense, and Attentional Representations for
  Text Retrieval}.
\newblock \bibinfo{journal}{\emph{Transactions of the Association for
  Computational Linguistics}}  \bibinfo{volume}{9} (\bibinfo{year}{2021}),
  \bibinfo{pages}{329--345}.
\newblock


\bibitem[\protect\citeauthoryear{Malkov and Yashunin}{Malkov and
  Yashunin}{2018}]%
        {malkov2018efficient}
\bibfield{author}{\bibinfo{person}{Yu~A Malkov} {and} \bibinfo{person}{Dmitry~A
  Yashunin}.} \bibinfo{year}{2018}\natexlab{}.
\newblock \showarticletitle{Efficient and robust approximate nearest neighbor
  search using hierarchical navigable small world graphs}.
\newblock \bibinfo{journal}{\emph{IEEE transactions on pattern analysis and
  machine intelligence}} \bibinfo{volume}{42}, \bibinfo{number}{4}
  (\bibinfo{year}{2018}), \bibinfo{pages}{824--836}.
\newblock


\bibitem[\protect\citeauthoryear{Muja and Lowe}{Muja and Lowe}{2009}]%
        {muja2009fast}
\bibfield{author}{\bibinfo{person}{Marius Muja} {and} \bibinfo{person}{David~G
  Lowe}.} \bibinfo{year}{2009}\natexlab{}.
\newblock \showarticletitle{Fast approximate nearest neighbors with automatic
  algorithm configuration.}
\newblock \bibinfo{journal}{\emph{VISAPP (1)}} \bibinfo{volume}{2},
  \bibinfo{number}{331-340} (\bibinfo{year}{2009}), \bibinfo{pages}{2}.
\newblock


\bibitem[\protect\citeauthoryear{Oord, Li, and Vinyals}{Oord
  et~al\mbox{.}}{2018}]%
        {oord2018representation}
\bibfield{author}{\bibinfo{person}{Aaron van~den Oord}, \bibinfo{person}{Yazhe
  Li}, {and} \bibinfo{person}{Oriol Vinyals}.} \bibinfo{year}{2018}\natexlab{}.
\newblock \showarticletitle{Representation learning with contrastive predictive
  coding}.
\newblock \bibinfo{journal}{\emph{arXiv preprint arXiv:1807.03748}}
  (\bibinfo{year}{2018}).
\newblock


\bibitem[\protect\citeauthoryear{Qu, Ding, Liu, Liu, Ren, Zhao, Dong, Wu, and
  Wang}{Qu et~al\mbox{.}}{2021}]%
        {qu2021rocketqa}
\bibfield{author}{\bibinfo{person}{Yingqi Qu}, \bibinfo{person}{Yuchen Ding},
  \bibinfo{person}{Jing Liu}, \bibinfo{person}{Kai Liu},
  \bibinfo{person}{Ruiyang Ren}, \bibinfo{person}{Wayne~Xin Zhao},
  \bibinfo{person}{Daxiang Dong}, \bibinfo{person}{Hua Wu}, {and}
  \bibinfo{person}{Haifeng Wang}.} \bibinfo{year}{2021}\natexlab{}.
\newblock \showarticletitle{RocketQA: An optimized training approach to dense
  passage retrieval for open-domain question answering}. In
  \bibinfo{booktitle}{\emph{Proceedings of the 2021 Conference of the North
  American Chapter of the Association for Computational Linguistics: Human
  Language Technologies}}. \bibinfo{pages}{5835--5847}.
\newblock


\bibitem[\protect\citeauthoryear{Ren, Zhang, and Li}{Ren et~al\mbox{.}}{2020}]%
        {ren2020hm}
\bibfield{author}{\bibinfo{person}{Jie Ren}, \bibinfo{person}{Minjia Zhang},
  {and} \bibinfo{person}{Dong Li}.} \bibinfo{year}{2020}\natexlab{}.
\newblock \showarticletitle{Hm-ann: Efficient billion-point nearest neighbor
  search on heterogeneous memory}.
\newblock \bibinfo{journal}{\emph{Advances in Neural Information Processing
  Systems}}  \bibinfo{volume}{33} (\bibinfo{year}{2020}).
\newblock


\bibitem[\protect\citeauthoryear{Shen, He, Gao, Deng, and Mesnil}{Shen
  et~al\mbox{.}}{2014}]%
        {shen2014learning}
\bibfield{author}{\bibinfo{person}{Yelong Shen}, \bibinfo{person}{Xiaodong He},
  \bibinfo{person}{Jianfeng Gao}, \bibinfo{person}{Li Deng}, {and}
  \bibinfo{person}{Gr{\'e}goire Mesnil}.} \bibinfo{year}{2014}\natexlab{}.
\newblock \showarticletitle{Learning semantic representations using
  convolutional neural networks for web search}. In
  \bibinfo{booktitle}{\emph{Proceedings of the 23rd international conference on
  world wide web}}. \bibinfo{pages}{373--374}.
\newblock


\bibitem[\protect\citeauthoryear{Singh, Subramanya, Krishnaswamy, and
  Simhadri}{Singh et~al\mbox{.}}{2021}]%
        {singh2021freshdiskann}
\bibfield{author}{\bibinfo{person}{Aditi Singh}, \bibinfo{person}{Suhas~Jayaram
  Subramanya}, \bibinfo{person}{Ravishankar Krishnaswamy}, {and}
  \bibinfo{person}{Harsha~Vardhan Simhadri}.} \bibinfo{year}{2021}\natexlab{}.
\newblock \showarticletitle{FreshDiskANN: A Fast and Accurate Graph-Based ANN
  Index for Streaming Similarity Search}.
\newblock \bibinfo{journal}{\emph{arXiv preprint arXiv:2105.09613}}
  (\bibinfo{year}{2021}).
\newblock


\bibitem[\protect\citeauthoryear{Subramanya, Kadekodi, Krishaswamy, and
  Simhadri}{Subramanya et~al\mbox{.}}{2019}]%
        {subramanya2019diskann}
\bibfield{author}{\bibinfo{person}{Suhas~Jayaram Subramanya},
  \bibinfo{person}{Rohan Kadekodi}, \bibinfo{person}{Ravishankar Krishaswamy},
  {and} \bibinfo{person}{Harsha~Vardhan Simhadri}.}
  \bibinfo{year}{2019}\natexlab{}.
\newblock \showarticletitle{Diskann: Fast accurate billion-point nearest
  neighbor search on a single node}. In \bibinfo{booktitle}{\emph{Proceedings
  of the 33rd International Conference on Neural Information Processing
  Systems}}. \bibinfo{pages}{13766--13776}.
\newblock


\bibitem[\protect\citeauthoryear{Xiao, Liu, Shao, Lian, and Xie}{Xiao
  et~al\mbox{.}}{2021}]%
        {xiao2021match}
\bibfield{author}{\bibinfo{person}{Shitao Xiao}, \bibinfo{person}{Zheng Liu},
  \bibinfo{person}{Yingxia Shao}, \bibinfo{person}{Defu Lian}, {and}
  \bibinfo{person}{Xing Xie}.} \bibinfo{year}{2021}\natexlab{}.
\newblock \showarticletitle{Matching-oriented Embedding Quantization For Ad-hoc
  Retrieval}. In \bibinfo{booktitle}{\emph{Proceedings of the 2021 Conference
  on Empirical Methods in Natural Language Processing}}.
  \bibinfo{publisher}{Association for Computational Linguistics},
  \bibinfo{pages}{8119--8129}.
\newblock
\urldef\tempurl%
\url{https://doi.org/10.18653/v1/2021.emnlp-main.640}
\showDOI{\tempurl}


\bibitem[\protect\citeauthoryear{Xiong, Xiong, Li, Tang, Liu, Bennett, Ahmed,
  and Overwijk}{Xiong et~al\mbox{.}}{2021}]%
        {xiong2021approximate}
\bibfield{author}{\bibinfo{person}{Lee Xiong}, \bibinfo{person}{Chenyan Xiong},
  \bibinfo{person}{Ye Li}, \bibinfo{person}{Kwok-Fung Tang},
  \bibinfo{person}{Jialin Liu}, \bibinfo{person}{Paul~N. Bennett},
  \bibinfo{person}{Junaid Ahmed}, {and} \bibinfo{person}{Arnold Overwijk}.}
  \bibinfo{year}{2021}\natexlab{}.
\newblock \showarticletitle{Approximate Nearest Neighbor Negative Contrastive
  Learning for Dense Text Retrieval}. In
  \bibinfo{booktitle}{\emph{International Conference on Learning
  Representations}}.
\newblock
\urldef\tempurl%
\url{https://openreview.net/forum?id=zeFrfgyZln}
\showURL{%
\tempurl}


\bibitem[\protect\citeauthoryear{Yang, Dai, Yang, Carbonell, Salakhutdinov, and
  Le}{Yang et~al\mbox{.}}{2019}]%
        {yang2019xlnet}
\bibfield{author}{\bibinfo{person}{Zhilin Yang}, \bibinfo{person}{Zihang Dai},
  \bibinfo{person}{Yiming Yang}, \bibinfo{person}{Jaime Carbonell},
  \bibinfo{person}{Russ~R Salakhutdinov}, {and} \bibinfo{person}{Quoc~V Le}.}
  \bibinfo{year}{2019}\natexlab{}.
\newblock \showarticletitle{Xlnet: Generalized autoregressive pretraining for
  language understanding}.
\newblock \bibinfo{journal}{\emph{Advances in neural information processing
  systems}}  \bibinfo{volume}{32} (\bibinfo{year}{2019}).
\newblock


\bibitem[\protect\citeauthoryear{Zhan, Mao, Liu, Guo, Zhang, and Ma}{Zhan
  et~al\mbox{.}}{2021}]%
        {zhan2021optimizing}
\bibfield{author}{\bibinfo{person}{Jingtao Zhan}, \bibinfo{person}{Jiaxin Mao},
  \bibinfo{person}{Yiqun Liu}, \bibinfo{person}{Jiafeng Guo},
  \bibinfo{person}{Min Zhang}, {and} \bibinfo{person}{Shaoping Ma}.}
  \bibinfo{year}{2021}\natexlab{}.
\newblock \showarticletitle{Optimizing Dense Retrieval Model Training with Hard
  Negatives}. In \bibinfo{booktitle}{\emph{Proceedings of the 44th
  International ACM SIGIR Conference on Research and Development in Information
  Retrieval}}. \bibinfo{pages}{1503–1512}.
\newblock


\bibitem[\protect\citeauthoryear{Zhang, He, Liu, Lim, and Bing}{Zhang
  et~al\mbox{.}}{2020}]%
        {zhang2020unsupervised}
\bibfield{author}{\bibinfo{person}{Yan Zhang}, \bibinfo{person}{Ruidan He},
  \bibinfo{person}{Zuozhu Liu}, \bibinfo{person}{Kwan~Hui Lim}, {and}
  \bibinfo{person}{Lidong Bing}.} \bibinfo{year}{2020}\natexlab{}.
\newblock \showarticletitle{An unsupervised sentence embedding method by mutual
  information maximization}.
\newblock \bibinfo{journal}{\emph{arXiv preprint arXiv:2009.12061}}
  (\bibinfo{year}{2020}).
\newblock


\end{thebibliography}

\appendix
\section{Algorithm Details}
\subsection{Notation Table}

\begin{table}[h!]
    \centering
    \small
    \begin{tabular}{p{2.5cm}p{5.0cm}}
        \hline
        Notation & Description  \\
        \hline
        $A$ & entire corpus \\
        $\widehat{{A}}_q$ & candidate set for query $q$ \\
        $\mathbf{C}$ & codebook for the quantization module \\
        $g(\cdot)$ &  query encoder in stage one \\
        $g'(\cdot)$ &  query encoder in stage two \\
        $g''(\cdot)$ & unified query encoder \\
        $f_s(\cdot)$ & sparse answer encoder \\
        $f_d(\cdot)$ & dense answer encoder \\
        $l(\cdot)$ & InfoNCE loss function \\
        \hline
    \end{tabular}
    \caption{Notation Table}
    \vspace{-20pt}
    \label{tab:notation}
\end{table}

\subsection{Locality Centric Sampling}
The pseudo-code is showed in Algorithm~\ref{alg:pseudocode}:
\begin{algorithm}[h!]
\begin{lstlisting}
# query2keys: 
#   for each query, storage its positive answers 
# query2negs: 
#   queries' adjacency list for bipartite proximity graph 
# neg2queries: 
#   answers' adjacency list for bipartite proximity graph 
# queries_set: set of unvisited queries
# random_sample: randomly sample one item

def get_next_query(q_queue, queries_set, strategy):
    # obtain the next query from queue
    if len(q_queue) == 0:
        return random_sample(queries_set)
    if strategy == 'RandomWalk':
        return q_queue[-1]
    if strategy == 'SnowBall':
        return q_queue[0]

def generate_batch_data(queries_set, 
                        query2keys, 
                        query2negs, 
                        neg2queries, 
                        strategy):
    # generate data for mini-batch training
    while len(queries_set) > 0:
        batch_data = [] 
        while len(batch_data) < batch_size:
              q = get_next_query(q_queue, queries_set, strategy)
            k = random_sample(query2keys[q])
            n = random_sample(query2negs[q])
            batch_data.append([q, k, n])
            
            q_queue.extend(neg2queries[n])
            # delete q from set of unvisited queries 
            queries_set.remove(q)
            # exit when all queries have be processed
            if len(queries_set) == 0: break
        return batch_data
\end{lstlisting}
\caption{Pseudocode of Locality Centric Sampling}\label{alg:pseudocode}
\end{algorithm}

\section{More Experiments Results}
More results are provided in additional to the reported values in Table~\ref{tab:1} and Table~\ref{tab:3}. The observations are consistent with our conclusion in Section~\ref{sec:analysis}.

\subsection{Evaluation of massive-scale EBR}
\begin{table}[h]
    \centering
    \footnotesize
    \begin{tabular}{p{1.8cm} | C{0.7cm} C{0.7cm} | C{0.7cm} C{0.7cm} | C{0.7cm} C{0.7cm} }
    \ChangeRT{1pt} &
    \multicolumn{2}{c|}{\textbf{Web Search(P)}} & \multicolumn{2}{c|}{\textbf{Web Search(D)}} & \multicolumn{2}{c}{\textbf{Sponsor Search}}   \\
    \cmidrule(lr){1-1}
    \cmidrule(lr){2-3}
    \cmidrule(lr){4-5}
    \cmidrule(lr){6-7}
    \textbf{Methods} & 
    \textbf{R@20} & \textbf{R@30} & \textbf{R@20} & \textbf{R@30} & \textbf{R@20} & \textbf{R@30} \\
    \hline
    DPR + DISK & 0.6373 & 0.6937 & 0.7038 &0.7529 &0.4086 &0.4575\\
    STAR + DISK  & 0.6767 & 0.7289 &0.7775 &0.8236 &0.4404 &0.4919 \\
    ANCE + DISK & 0.6744 & 0.7252 &0.7567 &0.7962 &0.4170 &0.4659 \\
    DE\_BERT + DISK &  0.6574 & 0.7102 &0.7090 &0.7600 &0.4233 &0.4682 \\
    \hline
    \textbf{BiDR(RD)} &0.7098 &\textbf{0.7604} & \textbf{0.7964} & \textbf{0.8399} &0.5160 &0.5776\\
    \textbf{BiDR(SN)} &\textbf{0.7116} &\textbf{0.7604} &0.7924 &0.8357 &\textbf{0.5217} &\textbf{0.5821}\\
    \hline
    \ChangeRT{1pt}
    \end{tabular}
    \caption{\small More results on massive-scale EBR in addition to Table~\ref{tab:1}.}
    \vspace{-5pt}
    \label{tab:appendix_1}
\end{table}

\subsection{Ablation study}
\begin{table}[h]
    \centering
    \footnotesize
    \begin{tabular}{p{1.8cm} | C{0.7cm} C{0.7cm} | C{0.7cm} C{0.7cm} | C{0.7cm} C{0.7cm} }
    \ChangeRT{1pt} &
    \multicolumn{2}{c|}{\textbf{Web Search(P)}} & \multicolumn{2}{c|}{\textbf{Web Search(D)}} & \multicolumn{2}{c}{\textbf{Sponsor Search}}   \\
    \cmidrule(lr){1-1}
    \cmidrule(lr){2-3}
    \cmidrule(lr){4-5}
    \cmidrule(lr){6-7}
    \textbf{Sampling} & 
    \textbf{R@20} & \textbf{R@30} & \textbf{R@20} & \textbf{R@30} & \textbf{R@20} & \textbf{R@30} \\
    \hline
    ANN &0.7048	&0.7553 &0.7845	&0.8304 &0.4998 &0.5604\\
    ANN + In-batch &0.7017 &0.7513 &0.7814 &0.8291 &0.4903 &0.5557\\
    Random Walk &0.7098 &\textbf{0.7604} & \textbf{0.7964} & \textbf{0.8399} &0.5160 &0.5776\\
    Snowball &\textbf{0.7116} &\textbf{0.7604} &0.7924 &0.8357 &\textbf{0.5217} &\textbf{0.5821}\\
    \hline \hline
    \textbf{Unified Query} & 
    \textbf{R@20} & \textbf{R@30} & \textbf{R@20} & \textbf{R@30} & \textbf{R@20} & \textbf{R@30} \\
    \hline
    Unified (RD) & 0.7008 &0.7536 &0.7927 &0.8368 &0.5072 &0.5664\\
    w.o. Unified (RD)  &0.7098 &\textbf{0.7604} & \textbf{0.7964} & \textbf{0.8399} &0.5160 &0.5776  \\
    Unified (SN) &0.7031 &0.7536 &0.7922 &0.8367 &0.5113 &0.5765\\
    w.o. Unified (SN) &\textbf{0.7116} &\textbf{0.7604} &0.7924 &0.8357 &\textbf{0.5217} &\textbf{0.5821}\\
    \ChangeRT{1pt}
    \end{tabular}
    \caption{\small More results on ablation studies in addition to Table~\ref{tab:3}.}
    \vspace{-5pt}
    \label{tab:appendix_2}
\end{table}

\section{Negative sampling in Stage One}
Despite that the ANN negative samples are useful in learning dense embeddings, we find that it does not contribute to the overall retrieval accuracy when applied to stage one (i.e., sparse embeddings' learning), as claimed in  Section~\ref{sec:negative_stage1}. Such a point is reflected by the following experimental results; i.e., ``w.o. ANN'' constantly leads to better performances.


\begin{table}[h]
    \centering
    \footnotesize
    \begin{tabular}{p{0.9cm} | p{0.7cm} | C{0.7cm} C{0.7cm} C{0.8cm} | C{0.7cm}  C{0.7cm} C{0.8cm} }
  \ChangeRT{1pt} \multicolumn{2}{c|}{\textbf{Sampling Method}} &
    \multicolumn{3}{c|}{\textbf{Web Search(P)}} & \multicolumn{3}{c}{\textbf{Web Search(D)}}    \\
    \cmidrule(lr){1-2}
    \cmidrule(lr){3-5}
    \cmidrule(lr){6-8}
    Sparse & Dense & 
    \textbf{R@10} & \textbf{R@50} & \textbf{R@100} & \textbf{R@10} & \textbf{R@50} & \textbf{R@100} \\
    \hline
    w.ANN & RW &0.6021 &0.8115 &0.8710 & 0.6880 &0.8678 &0.9079 \\
    w.o.ANN &RW & \textbf{0.6048} & \textbf{0.8156} & \textbf{0.8753} & \textbf{0.6988} & \textbf{0.8821} & \textbf{0.9176}\\
    \hline
    w.ANN &SN &0.5966 &0.8065 &0.8659 & 0.6832 &0.8740 &0.9110 \\
    w.o.ANN &SN & \textbf{0.6087} & \textbf{0.8169} & \textbf{0.8786} & \textbf{0.6959} & \textbf{0.8808} & \textbf{0.9225} \\
    \ChangeRT{1pt}
    \end{tabular}
    \caption{\small Impact of ANN negatives in sparse embedding learning.}
    \vspace{-5pt}
    \label{tab:hardneg_stage1}
\end{table}

\section{Implementation Details}
\subsection{System Specifications}
We conduct all experiments on
one machine with eight NVIDIA-A100-40G GPUs and two AMD EPYC 7V12 64-Core CPUs. 
The models are implemented with python 3.6 and pytorch 1.8.0. We also use FAISS\footnote{https://github.com/facebookresearch/faiss} in the implementation. 

\subsection{Text Encoder}
We use RoBERTa-base \cite{liu2019roberta} as the backbone of text encoder for all the methods. 
The output embedding of the “[CLS]” token is used for the text representation result. For Sponsor Search dataset, we reduce the embedding size from 768 to 64 by learning an additional linear projection layer.

\subsection{Hyper-parameter}
We use AdamW optimizer to update models' parameters. 
For Web Search(D) dataset, the batch size is 160, and the documents are truncated to a maximum of 512 tokens.
For Web Search(P), the batch size is set to be 1024 and the maximal length of documents is 128. For Sponsor Search, we set the maximal length of ads as 12, which allows us to use a larger batch size: 2048. 
The learning rates for Web Search(D), Web Search(P) and Sponsor Search are 5e-5, 1e-4, and 2e-4 respectively.

For the sparse embedding learning, the number of codewords in each codebook is fixed to be 256. 
We use OPQ to initialize the PQ module and the default number of codebooks is 96 on Web Search(P) and Web Search(D) datasets. Since the embedding size is smaller, we only use 8 codebooks for Sponsor Search.
For dense embedding learning, the top-200 irrelevant documents ranked by sparse embedding are used as the hard negative set to construct the bipartite proximity graph. 
Following \cite{xiong2021approximate} and \cite{zhan2021optimizing}, we use DE-BERT as the warm-up model and random sample one hard negative from top-200 documents for ANCE and STAR models.

\end{document}